\documentclass[a4paper, 11pt, DIV=12, abstract=true]{scrartcl}

\usepackage[backend=biber, style=numeric-comp, sorting=none]{biblatex}
\usepackage{authblk}

\usepackage{graphicx}
\usepackage{amsmath,amssymb,amsfonts}
\usepackage[title]{appendix}
\usepackage{xcolor}
\usepackage{hyperref}
\usepackage{listings}
\usepackage{subcaption}
\usepackage{algorithm}
\usepackage{algorithmicx}
\usepackage{algpseudocode}
\usepackage{enumitem}

\makeatletter
\newcommand{\email}[1]{%
  \unskip\thanks{\href{mailto:#1}{#1}}%
}
\makeatother

\newcommand*{\dphant}{{\ensuremath{\vphantom{\dagger}}}}

\newcommand{\mufix}{\text{FCP}}
\newcommand{\hfgen}{\text{GHF}}
\newcommand{\hfcnv}{\text{CHF}}

\begin{document}

\title{Phase stability, charge ordering, and charge-liquid formation in the Falicov-Kimball model on triangular and kagome lattices}
\date{}

\author[1,\email{a.nejati@fz-juelich.de}]{Ammar Nejati}

\author[2,3,\email{mlenk@uni-bonn.de}]{Marvin Lenk}

\affil[1]{Scientific Computing, Forschungszentrum J\"ulich, J\"{u}lich Centre for Neutron Science (JCNS), Lichtenbergstr. 1, 85747 Garching, Germany}
\affil[2]{Physikalisches Institut, Universit\"at Bonn, Nussallee 12, 53115 Bonn, Germany}
\affil[3]{\textit{Present affiliation:} Independent Researcher, Bonn, Germany}

\makeatletter
     \maketitle
     \normalfont
     \vspace{-4em}
     \begin{abstract}
     We present a systematic investigation of the stability of the phase diagram of the spinless Falicov-Kimball model on two-dimensional non-bipartite lattices, the triangular and kagome lattices, using Markov-chain Monte Carlo simulations. Three filling or chemical potential conditions are examined: fixed chemical potentials ($\mu_f = \mu_c = U/2$), ``generalized'' half-filling ($\bar{n}_f = 1/3$, $\bar{n}_c = 2/3$), and conventional half-filling ($\bar{n}_f = \bar{n}_c = 1/2$). For each condition, the effects of a perturbatively small next-nearest-neighbor hopping are studied. On the triangular lattice, Anderson insulator, Mott insulator, and charge-density wave (CDW) phases are found in all cases, with the CDW driven by Coulomb-interaction-mediated nesting. Evidence for a charge-liquid regime is found in the region between the weakly correlated regime and the CDW, characterized by competing charge-ordering wavevectors and the absence of universal scaling behavior, whose extent shrinks progressively from the grand-canonical to the conventional half-filling condition. On the kagome lattice, the CDW phase is completely absent; instead, an insulating ground state, plausibly of the Mott type, is found, with signatures of a possible quantum phase transition at zero temperature. Our results establish the crucial role of lattice geometry and particle-hole symmetry breaking in shaping the phase diagram of correlated electron systems.   
     \end{abstract}
     \vspace{2em}
\makeatother

\section{Introduction}

The Falicov-Kimball model (FKM) originated as a simplification of the Hubbard model, first conceived by Hubbard and Gutzwiller in the early 1960s~\cite{hubbard_1963_narrow_bands_1, hubbard_1964_narrow_bands_2, hubbard_1964_narrow_bands_3, gutzwiller_1965_narrow_band}. In 1969, Falicov and Kimball independently introduced additional features to study metal-insulator transitions in rare-earth and transition-metal compounds~\cite{falicov_kimball_1969, falicov_kimball_1970}, motivated by experimental observations of interactions between itinerant and localized electrons in these materials, particularly those involving valence fluctuations such as SmB$_{6}$ and YbInCu$_{4}$. The conventional FKM describes a lattice system with two flavors of \emph{spinless} fermions: itinerant (``mobile'') c-electrons with a dispersive band and localized (``frozen'') f-electrons with a narrow or flat band\cite{fan_two-dimensional_2023}. There is no direct interaction among f- or c-electrons themselves. Instead, the key ingredient of the model is an on-site Coulomb repulsion or attraction between these two species.

The simplicity of the Falicov-Kimball model, aptly designated as ``the Ising model of strongly correlated quantum systems''~\cite{jedrzejewski_lemanski_2001_falicov}, has enabled its application to a wide range of strongly correlated systems as a minimal framework for understanding the consequences of electron correlations, and its exact solvability in certain limits, such as with dynamical mean-field theory (DMFT)~\cite{freericks_exact_2003} for $d \rightarrow \infty$, makes it a valuable benchmark for exploring many-body effects in correlated electron systems.
Hence, it has been used to study metal-insulator transitions~\cite{chung_freericks_1998_metal_insulator, michielsen_1994_metal_insulator}, mixed-valence compounds~\cite{ramirez_falicov_1971_mixed_valence}, binary alloys~\cite{freericks_falicov_1990_fkm_binary_alloy}, crystallization~\cite{kennedy_lieb_1986_exact}, phase separation~\cite{kennedy_1998_phase_separation, freericks_1999_phase_separation_segregation}, segregation~\cite{lemberger_1992_segregation, freericks_1999_phase_separation_segregation, freericks_2002_segregation}, commensurate~\cite{freericks_2003_cdw_commensurate} and incommensurate~\cite{freericks_1993_cdw_incommensurate, freericks_lemanski_2000_segregation_cdw} charge-ordered phases, flux phases in magnetic fields~\cite{gruber_1996_flux_phase, gruber_1997_flux_configurations}, and the interplay between quantum particles and quasi-classical fields~\cite{macris_lebowitz_1997_rigorous, kirchner_2019_classical_quantum_liquid}. More recently, advances in ultra-cold experimental techniques have paved the way for the realization of the FKM with mixtures of light and heavy atoms in optical lattices~\cite{ates_ziegler_2005_mix_fermionic_ultracold, iskin_freericks_2009_dmft_optical_lattice}.

The model garnered renewed interest in the mid-1980s when mathematical physicists established rigorous results for the ground state of the FKM on bipartite lattices~\cite{freericks_exact_2003}.
In particular, Kennedy and Lieb~\cite{kennedy_lieb_1986_exact} and, independently, Brandt and Schmidt~\cite{brandt_schmidt_1986_exact} demonstrated that for an arbitrary bipartite lattice\footnote{
A lattice $\Lambda$ is bipartite if there exist two disjoint sets of sites A and B such that $\Lambda = A \cup B$ and the hopping amplitude between two sites vanishes when \emph{both} sites, $i$ and $j$, belong to the same sublattice.}
with dimension $d \geq 2$, at half-filling for both species and for finite values of the on-site Coulomb interaction strength, the conventional FKM possesses an inhomogeneous ground state exhibiting long-range order with a checkerboard-like distribution of the f-particles. Furthermore, Brandt and Mielsch~\cite{brandt_mielsch_1989_large_dim} obtained the exact solution of the FKM in the $d \rightarrow \infty$ limit and confirmed the existence of a charge-density wave (CDW) ground state in this limit.
It is important to note that the CDW phase in the FKM is not produced by the common Peierls-Fr\"{o}hlich instability due to electron-phonon coupling, but solely through the strong correlation between the f- and c-electrons~\cite{zhu_2017_charge_density_wave_origin}.
The available rigorous results hold only for \emph{bipartite} lattices, or only for large values of the on-site interaction $U$~\cite{gruber_macris_1996_falicov_kimball_exact, macris_lebowitz_1997_rigorous, kennedy_haller_2001_periodic_ground_state, gruber_2006_falicov_kimball}.
Perturbative results for large $U$ indicate that, to first order in $1/U$, the FKM on any lattice can be mapped onto an antiferromagnetic Ising Hamiltonian. On a triangular lattice, this leads to a geometrically frustrated Ising model, although the frustration is lifted when higher-order perturbative terms are taken into account~\cite{gruber_1997_flux_configurations}.

The characterization of the complex low-temperature phase of the FKM remains an active area of research, as a multitude of phases emerge depending on the model parameters and lattice geometries~\cite{jedrzejewski_lemanski_2001_falicov}.
In particular, using a Markov-chain Monte Carlo method~\cite{janke_2008_monte_carlo} adapted to FKM~\cite{maska_2006_thermodynamics, zonda_2009_phase_transition_3d_falicov_kimball, zonda_2012_phase_away_half_filling, czajka_2006_fkm_triangular}, Antipov et al.~\cite{kirchner_2016_interaction_induced_localization}. have shown that for the FKM on a 2d bipartite square lattice, apart from a CDW and Mott insulator phase, an Anderson-insulating phase can arise solely from an on-site interaction, even in the absence of explicit disorder. In this scenario, the inhomogeneous configurations of the f-electrons act as an \emph{effective annealed} disorder\footnote{
In the case of quenched disorder, the disorder configuration is static and the free energy is
$F_{\text{quench}} = -T\, \left\langle \ln Z \right\rangle_{\text{disorder}}$,
where $\left\langle \cdots \right\rangle_{\text{disorder}}$ denotes the average over disorder
realizations: each realization defines a separate physical system; observables are to be computed for each realization and then averaged.
In the case of annealed disorder, the disorder configuration is itself a thermodynamic variable that
fluctuates and equilibrates jointly with the rest of the system.
The free energy is $F_{\text{anneal}} = -T \ln \left\langle Z \right\rangle_{\text{disorder}}$,
where the partition function is first summed over disorder configurations, weighted by the appropriate Boltzmann factor.}
for the c-electrons, leading to localization of their wave functions.
Later, Oliveira et al. considered the FKM on a two-dimensional triangular lattice at a ``generalized'' half-filling and demonstrated how the interplay of localized and itinerant degrees of freedom, on the hexagonal lattice geometry, can lead to the emergence of a charge-liquid phase in the low-temperature part of the phase space, for moderate on-site interactions, in the proximity of the CDW phase.
They divided the charge-liquid region by a crossover line that terminates at a quantum critical point (QCP) and separates a ``quantum liquid'' (QL) from a ``classical liquid'' (CL), the behavior of which could possibly be captured by an effective, classical finite-ranged Ising-like model of f-electrons incorporating geometrical frustration. They found that by increasing the interaction strength, the CL undergoes a first-order transition to a CDW phase. The hallmark of this charge-liquid phase is the absence of universal behavior (in contrast to the CDW phase), observed in the specific heat and the static f-charge susceptibility~\cite{kirchner_2019_classical_quantum_liquid}.
A fundamental property of the FKM, established rigorously within DMFT by Si et al.~\cite{si_1992_breakdown_fermi_liquid_falicov_kimball}, is that the c-electron subsystem is generically \emph{not} a Fermi liquid for any value of mean f-occupation ($ 0 < \bar{n}_f < 1 $). This non-Fermi-liquid character is intrinsic to the model and arises from the local U(1) symmetry that prevents hybridization between the c- and f-electrons, which effectively act as a bath of coherent scatterers for the itinerant electrons. The non-Fermi-liquid nature of the FKM is directly relevant to the interpretation of the charge-liquid phase.

The stability of the phase diagram is of vital importance in any many-body system; i.e., the question whether phases are susceptible to deviations from symmetries or perturbations of model parameters. In this work, we present a systematic investigation of the stability of the previously observed phase diagram of the FKM on the 2d triangular and kagome lattices by imposing conditions on chemical potentials, the filling factors, and the lattice structure. For this purpose, we calculate a comprehensive set of observables to investigate the nature of each phase, beyond the scope of the previous works. The objective of this study is to assess the \emph{robustness} of the previously established phases, when the ensemble, the filling ratios, and the Fermi-surface geometry are varied. Accordingly, we adopt diagnostic criteria consistent with those of these prior works and augment them with additional observables. We avoid determination of the order or universality class of the transitions between the phases.

We first apply the same condition as for the square lattice~\cite{kirchner_2016_interaction_induced_localization} to investigate the pure effect of lattice geometry (square vs.\ hexagonal lattice). We then impose the ``generalized''~\cite{kirchner_2019_classical_quantum_liquid} and the standard half-filling conditions to study the consequences of different filling ratios. It is expected that the phases remain stable with respect to small deviations from exact filling ratios, as it is often numerically difficult to impose the exact conditions across the whole parameter range. Note that these cases correspond to two ensembles: first, the grand-canonical ensemble with $\mu_{c,f}$ fixed, allowing the particle numbers to fluctuate freely, and second, the canonical ensemble, in which the fillings $\bar{n}_{c,f}$ are fixed.

We subsequently add, in each case, a perturbatively small next-nearest-neighbor hopping to probe whether the phases are susceptible to slight modifications of the Fermi surface geometry. Such changes to the Fermi surface are expected to be consequential for the phases, especially the CDW phase, which relies on a nesting mechanism~\cite{zhu_2017_charge_density_wave_origin}.

Finally, for each case, we add a basis to the triangular lattice to obtain the kagome lattice in order to investigate whether the phases are altered by the addition of a basis.
It is expected that the phases of the kagome lattice would be starkly different from those of the triangular lattice due to the modified band structure and the Fermi surface (two dispersive bands and one flat band). Here, the role of the c-electron dispersionless band vis-\`{a}-vis the flat band of the f-electrons is of particular interest.

In all the cases above, we are interested, in particular, in the stability and robustness of the charge-liquid phase and its extension in the phase diagram.
We conclude with the overall phase diagrams, highlight our main observations and physical insights, and suggest avenues for future research.

\section{Model}
\subsection{Hamiltonian}
We consider the Falicov-Kimball model of spinless charged fermions (in the following often called electrons) on a lattice within the tight-binding approximation, including nearest- and next-nearest-neighbor hopping. The Hamiltonian reads
\begin{align} \nonumber
H_{\mathrm{FKM}} &= -t \sum_{\langle i,j \rangle} \big( c_i^\dagger c_j^\dphant + \mathrm{h.c.} \big)
- t_{\text{NNN}} \sum_{\langle \langle i, j \rangle \rangle} \big( c_{i}^\dagger \, c_{j}^\dphant + \mathrm{h.c.} \big)\\
&\phantom{=}+ U \sum_i \hat{n}_{c,i} \, \hat{n}_{f,i}
- \sum_i ( \mu_c \, \hat{n}_{c,i} + \mu_f \,\hat{n}_{f,i}),
\end{align}
where $c^\dagger_i$ ($c^\dphant_i$) is the creation (annihilation) operator of itinerant spinless electrons at site $i$, $\hat{n}_{c,i}$ ($\hat{n}_{f,i}$) the respective number operator of itinerant (localized) electrons, $\langle i,j \rangle$ ($\langle\langle i,j \rangle\rangle$) denotes the sum over (next-) nearest neighbors, $t$ ($t_{\text{NNN}}$) is the (next-) nearest-neighbor hopping amplitude, $U$ is the local Coulomb repulsion between c- and f-electrons and $\mu_c$ and $\mu_f$ the respective chemical potentials.

To mitigate boundary effects inherent to finite-size simulations, we impose periodic boundary conditions.

\subsection{Lattice geometry}
\begin{figure}[t]
    \centering
    \begin{subfigure}[b]{0.45\columnwidth}
        \centering
        \includegraphics[width=\textwidth]{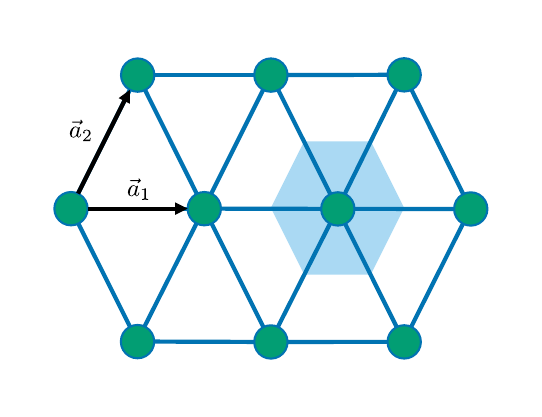}
        \label{fig:triangular_lattice}
    \end{subfigure}
    \hfill
    \begin{subfigure}[b]{0.45\columnwidth}
        \centering
        \includegraphics[width=\textwidth]{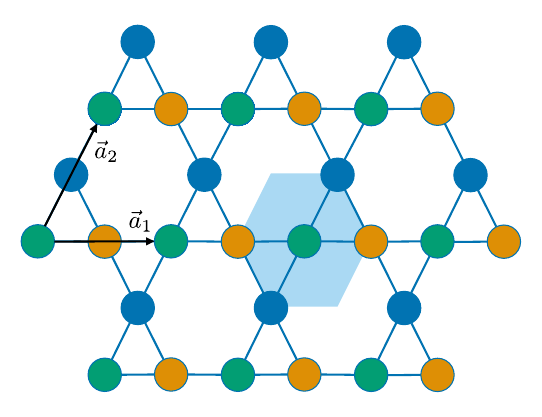}
        \label{fig:kagome_lattice}
    \end{subfigure}

    \caption{Illustration of finite-sized triangular (left) and kagome (right) lattices. The light-blue hexagon indicates the conventional unit cell.}
    \label{fig:lattices}
\end{figure}

We focus on the non-bipartite triangular and kagome lattices. A lattice is termed bipartite in the graph-theoretic sense if its sites can be partitioned into two disjoint sets such that all nearest-neighbor connections occur between, and not within, the sets.

The triangular lattice under consideration has a single-atom basis, where the real-space Bravais vectors are given by $ \mathbf{a}_1 = \{1, 0\} $, $ \mathbf{a}_2 = \frac{1}{2} \{ 1, \sqrt{3} \} $. It is illustrated in the left panel of Fig.~\ref{fig:lattices}. Conveniently, the reciprocal lattice itself forms a rescaled and rotated triangular lattice with lattice vectors $ \mathbf{b}_1 = \frac{4 \pi}{\sqrt{3}} \{ \frac{\sqrt{3}}{2}, -\frac{1}{2} \} $, $ \mathbf{b}_2 = \frac{4 \pi}{\sqrt{3}} \{ 0, 1 \} $. The Wigner-Seitz cell is therefore of hexagonal shape. The coordination number of this lattice is $Z = 6$, comprising three pairs of nearest neighbors related by the $D_6$ symmetry. Generally, this coordination number can lead to geometrical frustration of nearest-neighbor interactions such as antiferromagnetism and charge order at most fillings, permitting the emergence of exotic phases including spin liquids, unconventional superconductivity, and multiferroicity. Due to the spinless nature of the electrons considered here, the simplest expected ordered phase is a CDW.

The kagome lattice is a lattice with a three-atom basis of corner-sharing triangles, forming a triangular superlattice. It is illustrated in the right panel of Fig.~\ref{fig:lattices}, where the basis is indicated by color-coding the lattice points. This geometry yields a coordination number of $Z = 4$, while the Bravais superlattice maintains $Z_{\mathrm{Bravais}} = 6$. The tight-binding model of the kagome lattice features two dispersive bands that may form a Dirac cone at the band-touching points, enabling the possibility of massless high-mobility charge carriers and nontrivial topology. In addition, a single fully dispersionless (flat) band enhances correlations significantly due to its localized nature. This juxtaposition of massless Dirac carriers with flat-band heavy-fermion physics endows the kagome lattice with the potential to host a multitude of exotic phases beyond those of the simple triangular lattice.

\subsection{Particle-hole symmetry}
One of the fundamental symmetries of the FKM is the particle-hole (PH) symmetry.
The full PH symmetry transformation for a \emph{bipartite} lattice is defined
for the particle operators $ f $ and $ c $ as
\begin{align}
  f_{l \alpha} & \stackrel{\mathrm{PH}}{\mapsto} (-1)^{p(l)} \, {f^{h}_{l \alpha}}^\dagger , \\
  c_{l \alpha} & \stackrel{\mathrm{PH}}{\mapsto} (-1)^{p(l)} \, {c^h_{l \alpha}}^\dagger ,
\end{align}
where $ f^h $ and $ c^h $ are the hole operators and
$ p(l) = 0 $ on sublattice A, whereas $ p(l) = +1 $ for sublattice B.
The alternating sign factor $(-1)^{p(l)}$ is essential: it ensures that the hopping term, which connects opposite sublattices, changes sign under the transformation, thereby mapping the kinetic-energy operator onto itself.

Under the PH transformation, the Falicov-Kimball Hamiltonian
(in absence of next-nearest-neighbor hopping, $t_{\text{NNN}} = 0 $) maps to
\begin{align}\nonumber
  P H_{\mathrm{FKM}} P^\dagger &= H_{\mathrm{FKM}}
  - U (N_f + N_c - N_{\mathrm{site}})\\
  &\phantom{=}- (\mu_c + \mu_f) N_{\mathrm{site}}
  + 2 (\mu_f \, N_f + \mu_c \, N_c)\,.
\end{align}
On a bipartite lattice, the exact PH symmetry is achieved when the residual terms vanish.
This happens, in particular, in two special cases: $ \mu_f = \mu_c = U / 2 $, and $ \bar{n}_f = \bar{n}_c = \frac{1}{2} $.

On the non-bipartite triangular and kagome lattices, the full PH symmetry of the Hamiltonian is \emph{intrinsically broken} by the lattice geometry, as no sublattice decomposition exists that makes all hopping terms connect opposite sublattices.

Therefore, the conditions $\mu_f = \mu_c = U/2$ and $\bar{n}_f = \bar{n}_c = 1/2$ do not restore a true PH symmetry of the Hamiltonian. This intrinsic breaking of PH symmetry on non-bipartite lattices is one of the key factors that distinguishes the physics studied here from the well-understood bipartite case, and is expected to qualitatively affect the nature of the CDW ordering and the extent of the charge-liquid regime.

\subsection{Local \texorpdfstring{$U(1)$}{U(1)} symmetry}
A crucial property of the Falicov-Kimball model is the local $U(1)$ symmetry arising from the fact that the Hamiltonian commutes with the f-electron number operator at each site, $[\hat{H}_{\mathrm{FK}}, \hat{n}_{f, \alpha}(\mathbf{r}_i)] = 0$; hence, each $n_{f, \alpha}(\mathbf{r}_i)$ is individually conserved in the FKM, contrary to e.g.~Anderson impurity models.
This local symmetry, in fact, precludes ``spontaneous c-f hybridization'', which would lead to $\langle c^\dagger_i f_i \rangle \neq 0$ (or its Hermitian conjugate), at any finite temperature. The most significant physical consequence of this is that the hallmarks of the Anderson lattice model, like the mixed-valence regime and the Kondo-screened state (both of which rely on a coherent c-f hybridization), are absent in the FKM.

\section{Method}\label{sec:method}
\begin{algorithm*}[t]
\caption{Metropolis MCMC pseudo-code for the Falicov-Kimball model.}\label{alg:mcmc}
\begin{algorithmic}[1]
\Procedure{FKM-MCMC}{Lattice type, $T, U, \mu_{c,f}, t_\text{NNN}, N_\text{MC}, n_\text{skip}$}
\State Generate a random initial f-configuration $\{n_f\}$
\State Diagonalize $\mathcal{H}_c[\{n_f\}]$ to compute $\mathcal{W} \gets e^{\beta \mu_f \sum_{i\alpha} n_{f,\alpha}(\mathbf{r}_i)} \, \mathcal{Z}_c[\{n_f\}]$
\For{$s = 1, 2, \ldots, N_\text{MC}$}
  \State Select a random lattice site $(i, \alpha)$ \Comment{Propose}
  \State $\{n_f'\} \gets \{n_f\}$ with $n_{f,\alpha}(\mathbf{r}_i) \to 1 - n_{f,\alpha}(\mathbf{r}_i)$
  \State Diagonalize $\mathcal{H}_c[\{n_f'\}]$ to compute $\mathcal{W}' \gets \mathcal{W}[\{n_f'\}]$
  \State Accept the proposed $\{n_f\}$ with probability $P_{\mathrm{accept}} = \min\!\left(1,\, \mathcal{W}'/\mathcal{W}\right)$: \Comment{Metropolis}
  \Statex \{
  \State $r \gets \mathcal{W}' / \mathcal{W}$;\; draw $u \sim \mathrm{Uniform}[0,1)$
  \If{$u < r$}
    \State $\{n_f\} \gets \{n_f'\}$;\; $\mathcal{W} \gets \mathcal{W}'$ \Comment{Accept}
  \EndIf
  \Statex \}
  \If{$s \bmod n_\text{skip} = 0$} \Comment{Measure}
    \State Record basic observables using $\{n_f\}$ and eigenstates of $\mathcal{H}_c[\{n_f\}]$
  \EndIf
\EndFor
\State Compute thermal averages from recorded samples
\EndProcedure
\end{algorithmic}
\end{algorithm*}

\subsection{Monte Carlo simulation}
Utilizing the local $U(1)$ symmetry mentioned above, the partition function can be rewritten as a sum over classical f-configurations, $\{ n_f \}$,
\begin{align}
  \mathcal{Z}_{\mathrm{FK}} & = \sum_{ \{ n_f \} } e^{ \beta \mu_f \sum_{i \alpha} n_{f, \alpha}(\mathbf{r}_i) } \, \mathcal{Z}_c ~, \nonumber \\
  \mathcal{Z}_c & = \mathrm{Tr}_c \, e^{ - \beta \mathcal{H}_c[ \{ n_f \} ] } \,,
\end{align}
where $\mathcal{H}_c$ is an effective \emph{quadratic} Hamiltonian for the c-electrons for a given f-configuration, and $\mathcal{Z}_c$ is the corresponding partition function. The importance sampling is performed via the classical (sign-free) Markov-chain Monte Carlo method with the Metropolis algorithm~\cite{janke_2008_monte_carlo}.
With this, we simulate the Falicov-Kimball model (FKM) on finite lattices with periodic boundary conditions. The data used for this work are extracted from simulations with $ 2^{20} > 10^6 $ Monte Carlo (MC) steps with an ``auto-correlation time'' (skipped steps) of at most 20 steps.
A schematic description of the MCMC algorithm is given in Algorithm~\ref{alg:mcmc}.

The computational bottleneck is the diagonalization of $\mathcal{H}_c$ at each proposed step, which scales as $\mathcal{O}(N_\mathrm{site}^3)$. For the system sizes considered in this work, this remains tractable.

When $\bar{n}_{f,c}$ must be fixed to a given $\bar{n}_{f,c}^0$, the solutions to the coupled set of nonlinear equations for $\mu_{f,c}$, $ \{ \bar{n}_c(\mu_f, \mu_c) = \bar{n}_c^0,\; \bar{n}_f(\mu_f, \mu_c) = \bar{n}_f^0 \}$, must be obtained to be used as input to the main MC simulation.
To determine the solution to these equations, a separate preceding set of MC simulations is required; hence, the equations are stochastic due to the nature of the underlying MC simulations. In contrast to previous works~\cite{kirchner_2019_classical_quantum_liquid}, we employ the Nelder-Mead algorithm~\cite{nelder_simplex_1965}, a direct-search \emph{gradient-free} optimization method, to determine $\mu_{f,c}$, improving the numerical efficiency, precision, and confidence of the solution.

At low temperatures, standard implementations of this method can suffer from numerical instabilities. To mitigate these issues, we adapt the algorithm to maintain numerical stability as effectively as possible. The numerical results of the MCMC simulation are verified by comparing with the exact results for small lattice sizes.

A finite lattice possesses a discrete single-particle spectrum, and the associated mean level spacing sets a natural lower bound on the temperature range that can be resolved reliably. For the (non-interacting) tight-binding limit of triangular lattice, with bandwidth $W$ distributed over $N_{\mathrm{site}}$ states, a rough estimate of the level spacing is $\delta\varepsilon \sim W / N_{\mathrm{site}} $ ($\approx 0.06\,t $ for $ L = 12 $). This uniform-density estimate is, however, only an order-of-magnitude guide: once the Coulomb interaction is switched on, the spectrum is substantially reorganized into narrow bundles of levels separated by a pronounced correlation gap near the Fermi edge, so that the locally relevant level spacing is generically smaller, of order $\mathcal{O}(10^{-2}) \, t$. The temperatures analyzed in this work lie in the range $T \in [0.01, 0.15]$ (see Appendix~\ref{app:parameter_space}), i.e., predominantly of order $\mathcal{O}(10^{-1})\,t$, which is comparable to or larger than this spacing. Consequently, thermal broadening is at least of the order of the discretization scale throughout the bulk of the studied range, and the reported observables are physically meaningful. Only at the very lowest temperature, $T \sim 0.01 \, t$, does the level spacing become competitive with $T$; the associated caveats, together with the concomitant drop in the Markov-chain acceptance rate, are addressed explicitly in the Conclusion.

\subsection{Observables}
To characterize the phases, determine their nature, and identify the associated transitions, we compute a comprehensive set of physical observables listed below. We denote the temperature as $T$ and define the volume $V = L^2$ in terms of the linear system size $L$. The total number of lattice sites is $N_{\mathrm{site}} = n_{\mathrm{basis}} \, L^2$, with $n_{\mathrm{basis}}$ denoting the number of atoms per unit cell.

\begin{enumerate}
\item Average occupation densities $\bar{n}_{f,c} = N_{f,c} / N_{\mathrm{site}}$, where $N_{f,c}$ is the total number of c- or f-electrons,
  $ N_{f,c} = \sum_{i \in \text{lattice}} n_{f,c}(\mathbf{r}_i) $.

    \item ``Double-flavor'' occupation, $\langle n_f(\mathbf{r}_i) \; n_c(\mathbf{r}_i) \rangle$, which quantifies the probability of simultaneous occupation of a lattice site $\mathbf{r}_i$ by electrons of both species.

    \item Specific heat capacity at constant volume, $c_v(T) = \frac{1}{V} \, \mathrm{d}Q / \mathrm{d}T |_V$. In practice, $c_v(T)$ is evaluated directly
      from the fluctuations in the total internal energy $E$, $c_v(T) = \big( \langle E^2 \rangle - \langle E \rangle^2 \big) / (V T^2)$. A local maximum in the specific heat signals a transition (phase transition or crossover). The CDW phase border, for each $ U $, is determined by the temperature at which the specific heat attains its maximum.
We have verified that peaks in $c_v(T)$ arising from the bare tight-binding band structure (at $U \rightarrow 0$) lie outside the studied temperature range; hence, all observed peaks can be attributed solely to the Coulomb interaction.

    \item Inverse participation ratio (IPR), as a dimensionless measure of eigenstate localization~\cite{wegner_inverse_1980,evers_anderson_2008}:
    \begin{align}
    \mathrm{IPR}(\varepsilon_\lambda) &= \sum_{i \in \text{lattice}} \left( \frac{| \langle \mathbf{r}_i | \varphi_\lambda \rangle |^2 }
         { \sum_{l \in \text{lattice}} | \langle \mathbf{r}_l | \varphi_\lambda \rangle |^2 } \right)^2 \,,
    \end{align}
    where $\varphi_\lambda$ is an eigenstate of $\mathcal{H}_c$ with eigenvalue $\varepsilon_\lambda$. The IPR near the Fermi edge attains a minimum value of $1 / N_{\mathrm{site}} \sim L^{-d} $ for a completely delocalized state and approaches $ \xi^{-d} $ for a fully localized state, with $ \xi $ being the localization length.
    IPR approaches its maximal value, within the possible range, in the Anderson insulator (AI) and Mott insulator (MI) phases. 

    \item The c-electron density of states (c-DoS), $\rho_{c} ( \varepsilon, T ) = \frac{1}{V} \sum_{k} \delta(\varepsilon - \varepsilon_{k})$, and its value at the Fermi edge, $ \rho_c(\varepsilon = E_F) $, are used to distinguish between the insulating Anderson and Mott phases.

    \item The gap around the Fermi edge, $ \Delta(E_F, T) $.

    \item The \emph{classical} f- and c-electron charge susceptibilities, $ \chi_{\alpha}(\mathbf{q}, T) $, defined as:
    \begin{gather}
        \chi_{\alpha}(\mathbf{r}_i, \mathbf{r}_j; T) = \langle n_{\alpha} (\mathbf{r}_i) \, n_{\alpha} (\mathbf{r}_j) \rangle
          - \langle n_{\alpha} (\mathbf{r}_i) \rangle \langle n_{\alpha} (\mathbf{r}_j) \rangle\,, \nonumber \\
        \chi_{\alpha}(\mathbf{q}, T) = \frac{\beta}{V} \sum_{i, j \in \text{lattice}}
        e^{- i \, \mathbf{q} \cdot (\mathbf{r}_i - \mathbf{r}_j) } \, \chi_{\alpha}(\mathbf{r}_i, \mathbf{r}_j; T) \,,
    \end{gather}
    where $\chi_{\alpha}$ denotes the f- or c-susceptibility, with $\alpha$ labeling the Wannier orbital (basis atom) in the case of the kagome lattice. Note that this \emph{classical} susceptibility is constructed from density expectation values, $n_{\alpha} (\mathbf{r}_i)$, in contrast to the \emph{quantum} susceptibility which directly involves density operators, $\hat{n}_{\alpha} (\mathbf{r}_i)$.

    \item Binder cumulants (4th-order cumulants) for the f- and c-species, $B_{f,c}(\mathbf{q}, T)$, which serve to classify the nature of phase transitions:~\cite{binder_critical_1981,binder_finite_1981,binder_finite-size_1984}
    \begin{equation}
     B_{\alpha}(\mathbf{q}, T) = 1 - \frac{ \langle | n_{\alpha}(\mathbf{q}, T) |^4 \rangle }
        {3 \left( \langle | n_{\alpha}(\mathbf{q}, T) |^2 \rangle \right)^2 }\,,
    \end{equation}
    where $ n_{\alpha, \mathbf{q}}(T) = \sum_{i \in \text{lattice}} e^{- i \, \mathbf{q} \cdot \mathbf{r}_i } \, n_{\alpha}(\mathbf{r}_i, T) $ is the Fourier transform of the density of the $ \alpha $-basis at site $ i $.
\end{enumerate}

To obtain c-DoS and IPR as continuous functions of energy, $ \omega $, a simple smoothing procedure is applied. The smoothed $ \rho_c(\omega) $ is always normalized to unity (see Appendix~\ref{app:smoothing}).

\subsection{Determining phases and borders}
Three phases are expected to appear in the low-temperature regime of the FKM: Mott insulator (MI), Anderson insulator (AI), and charge-density wave (CDW). They are identified and distinguished by the following criteria:
\begin{itemize}
    \item[MI:] Maximal IPR at $E_F$, $\rho_{c} (E_F, T) \rightarrow 0$, gap $ \Delta(E_F) > 0$, suppressed double-flavor occupation.
    \item[AI:] Maximal IPR at $E_F$, $\rho_{c} (E_F, T) > 0$.
    \item[CDW:] Small IPR at $E_F$, finite gap $ \Delta(E_F) $ at the Fermi level, modulated density. Additionally, sharp peak in $c_v$, simultaneous divergence in $\chi_{c,f}$, and change of Binder-cumulant behavior around phase-transition temperature.
\end{itemize}

The crossover between AI and MI phases is determined by a threshold in $ \rho_c(E_F) $: When $ \rho_c(E_F) $ drops below 0.01 of its value at the initial $U$-value, then an MI phase is assumed, as the states at the Fermi edge are depleted and a Mott gap is formed due to the increasing Coulomb-interaction strength.


The CDW-instability exhibits a twofold character in that it may or may not display a universal scaling behavior. We determine the regions of ``universality'' by the following criteria:
\begin{enumerate}[label=\roman*]
    \item The f- and c-electron susceptibilities reach a maximum at the \emph{same} temperature.
    \item The maximum value of the f- and c-electron susceptibilities increases with increasing $L$.
    \item  The f- and c-electron Binder cumulants exhibit the characteristic behavior of \emph{continuous} phase transitions.
\end{enumerate}
A violation of any single criterion indicates non-universal behavior.
We stress that a failure of universal scaling is, in itself, not sufficient to establish a distinct thermodynamic phase: it is equally the expected phenomenology of a \emph{first-order} transition, which for the FKM in the infinite-coordination limit is known to be generic and to be accompanied by higher-period ordered phases~\cite{gruber_2001_higher_period}.
Where the CDW universality criteria are violated, we identify a \emph{charge-liquid} regime (the terminology introduced by Ref.~\cite{kirchner_2019_classical_quantum_liquid}), a correlated, unordered region of the phase diagram intermediate between the weakly correlated regime and the CDW phase; on the finite systems, it is most accurately characterized as a strongly frustrated finite-temperature crossover, in which several charge-ordering wavevectors compete and no long-range CDW order is established, rather than as a sharply defined thermodynamic phase.
We note that these signatures provide operational evidence for a correlated non-Fermi-liquid regime, consistent with the intrinsic non-Fermi-liquid character of the FKM.~\cite{si_1992_breakdown_fermi_liquid_falicov_kimball}

\section{Results and discussion}
The results are organized as follows. We consider the FKM with three distinct chemical potential or filling conditions:
\begin{enumerate}
    \item[\mufix{}:] Fixed chemical potential $\mu_c = \mu_f = U/2$.
    \item[\hfgen{}:] Generalized half-filling $\bar{n}_c = 2/3$, $\bar{n}_f = 1/3$.
    \item[\hfcnv{}:] Conventional half-filling $\bar{n}_c = \bar{n}_f = 1/2$.
\end{enumerate}

The parameter range for the phase-space scan is decided with the computational cost in mind. Along with major cost of diagonalization, the determination of the chemical potentials for the cases where the average particle numbers are fixed is computationally very expensive. For a detailed exposition of the investigated parameter space, see Appendix~\ref{app:parameter_space}.

The strength of the nearest-neighbor hopping, $ t $, is always fixed at 1.0 and used as a scale for other
parameters; hence, the reported numerical values for $ T $, $ U $ and energies are always in units of $ t $. Natural units are always assumed, $ \hbar = k_B = c = 1 $ and lattice constant $ a = 1 $.

We first analyze the FKM on a triangular lattice with next-nearest-neighbor hopping of various strengths ($t_{\text{NNN}} / t \in [0, 0.1]$). We then add a basis to the nearest-neighbor triangular-lattice tight-binding model (with $ t_{\text{NNN}} = 0 $) to obtain the kagome lattice.

\subsection{Triangular lattice}
\begin{figure*}[!t]
    \centering
    \begin{subfigure}[b]{0.325\textwidth}
        \centering
        \includegraphics[width=\textwidth]{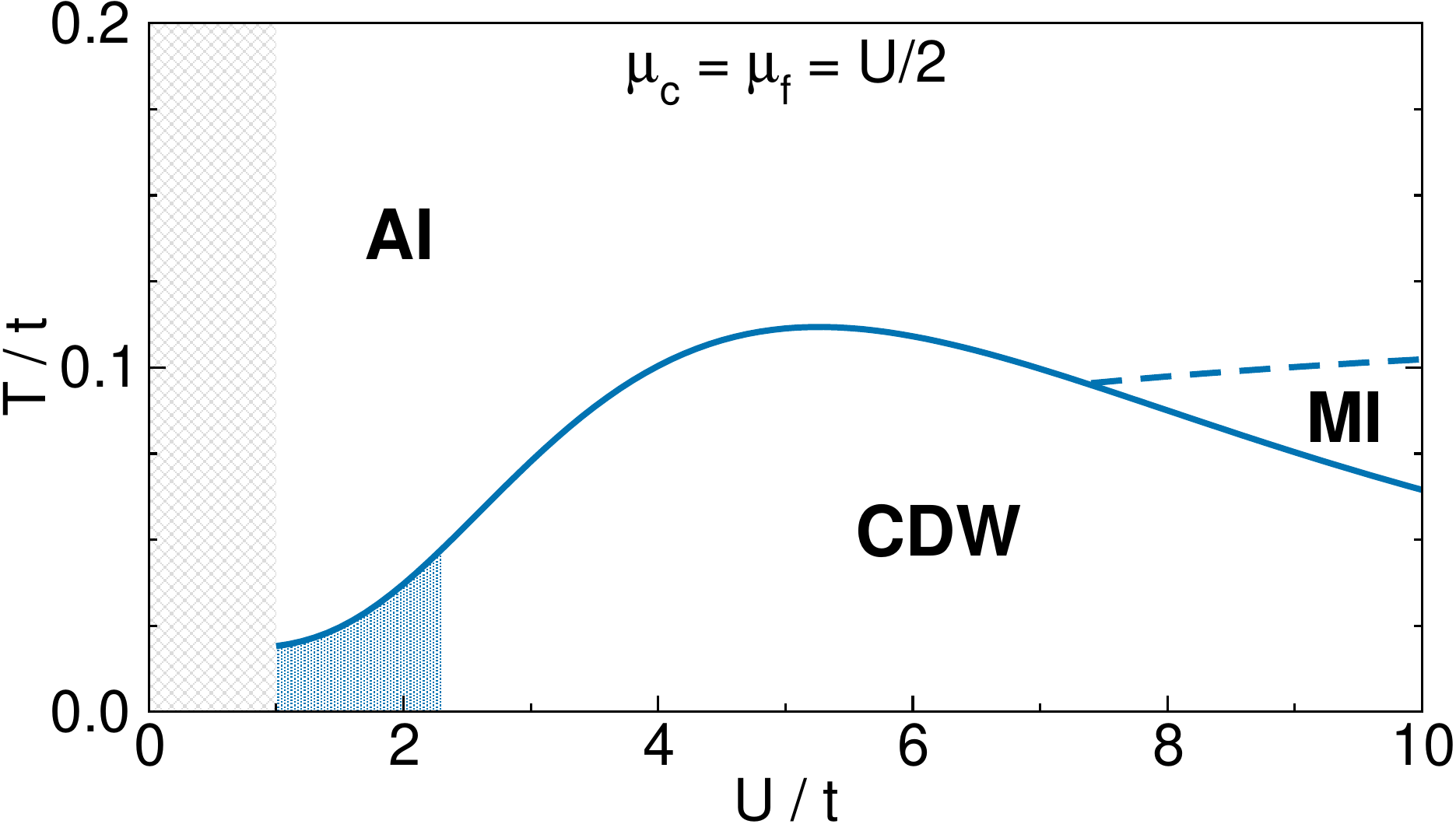}
        \label{fig:phase_diagram:N0}
    \end{subfigure}%
    \hfill
    \begin{subfigure}[b]{0.325\textwidth}
        \centering
        \includegraphics[width=\textwidth]{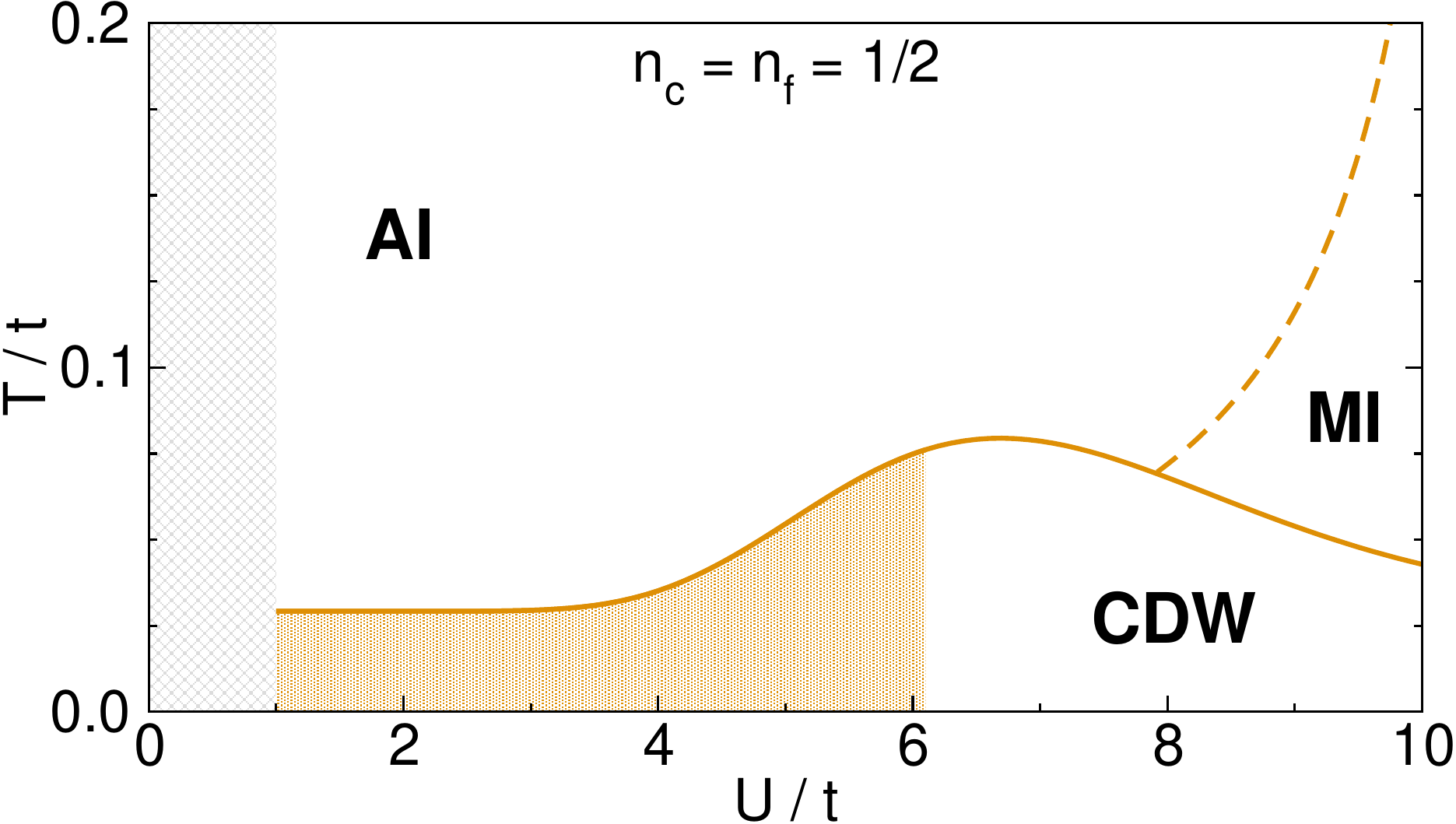}
        \label{fig:phase_diagram:N1}
    \end{subfigure}
    \hfill
    \begin{subfigure}[b]{0.325\textwidth}
        \centering
        \includegraphics[width=\textwidth]{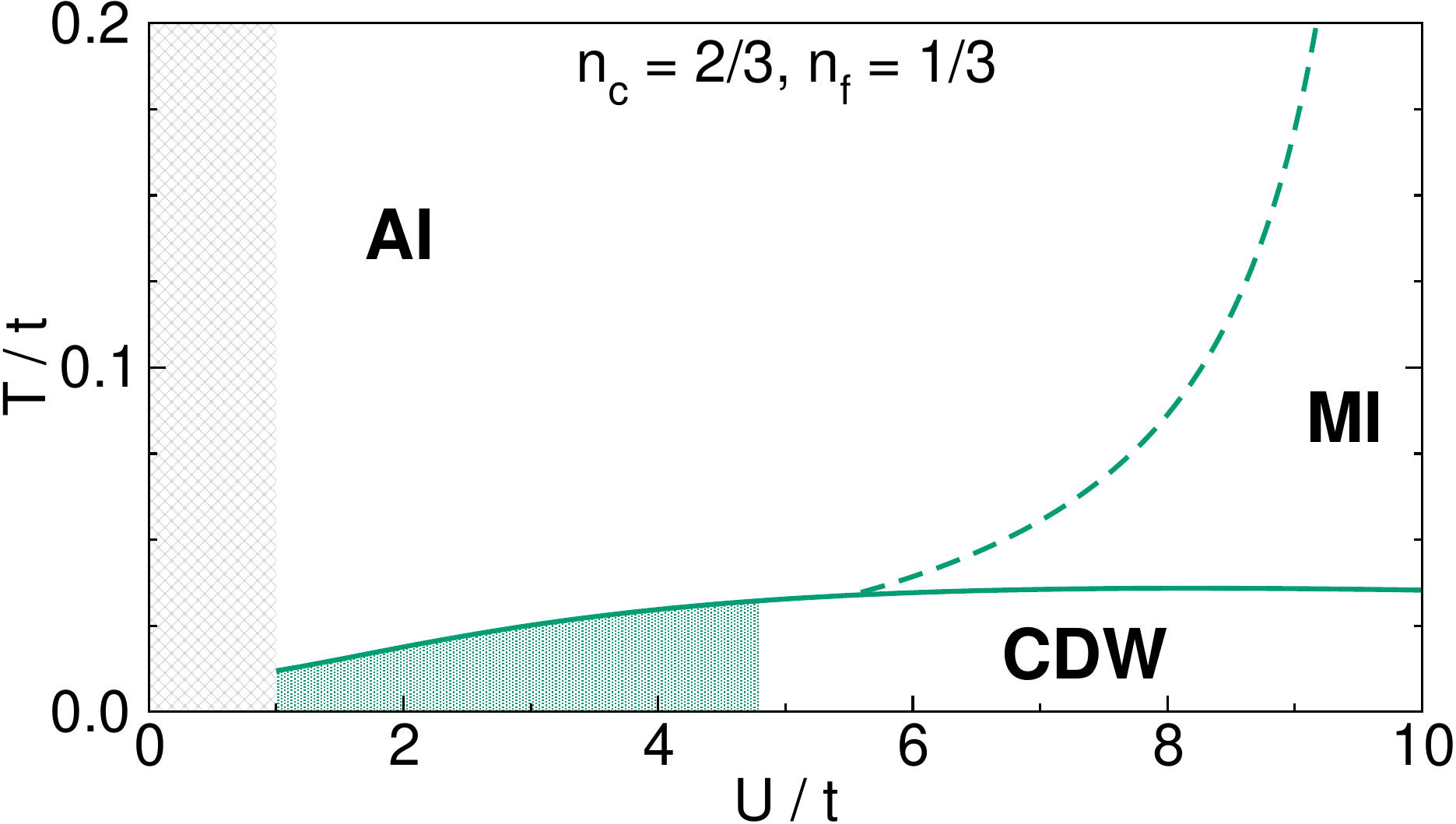}
        \label{fig:phase_diagram:N2}
    \end{subfigure}

    \caption{Phase diagram of the triangular lattice for \mufix{}, \hfcnv{}, and \hfgen{} cases as mentioned in the text, at vanishing next-nearest-neighbor hopping, $t_{\text{NNN}} = 0$. The region $U / t < 1$ is grayed out due to restrictions of the applied methods. The colored shaded regions mark the charge-liquid sector, i.e.\ the non-universal regime that precedes the onset of long-range CDW order, where several ordering wavevectors compete with each other.}
    \label{fig:phase_diagrams}
\end{figure*}
The phase diagrams for the three cases are presented in Fig.~\ref{fig:phase_diagrams}. All three cases host AI, MI, and CDW phases in the low-temperature regime. Notably, in the \mufix{} case, the CDW phase persists to higher temperatures and the AI phase extends to larger interaction strengths than in the \hfgen{} and \hfcnv{} cases.

\begin{figure}[tbh]
    \centering
    \begin{subfigure}{0.47\columnwidth}
        \includegraphics[width=\textwidth]{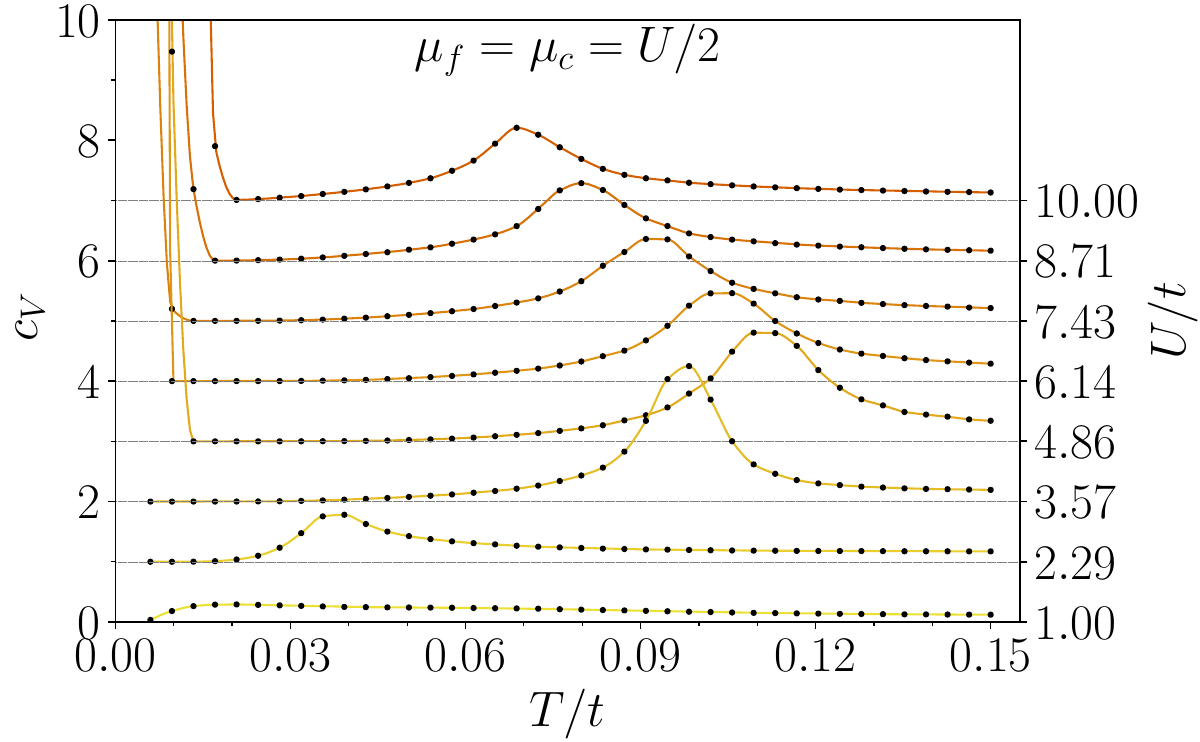}
    \end{subfigure}
    \hfill
    \begin{subfigure}{0.47\columnwidth}
        \includegraphics[width=\textwidth]{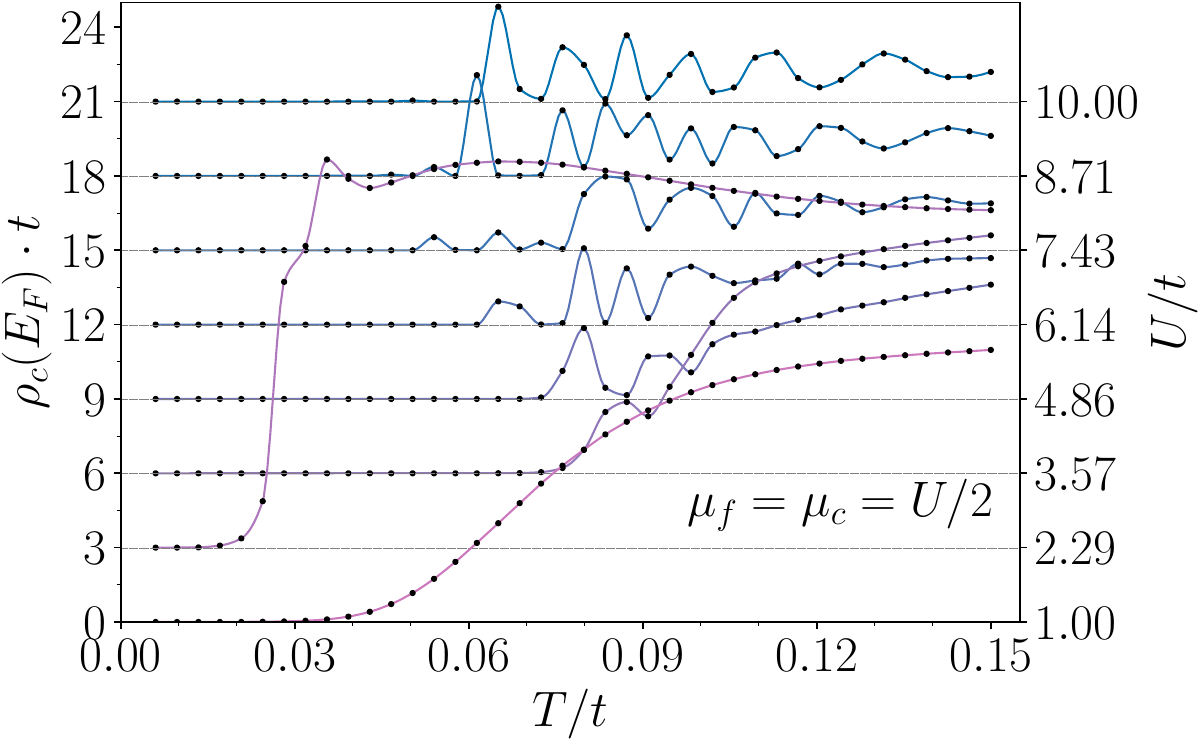}
    \end{subfigure}

    \begin{subfigure}{0.47\columnwidth}
        \includegraphics[width=\textwidth]{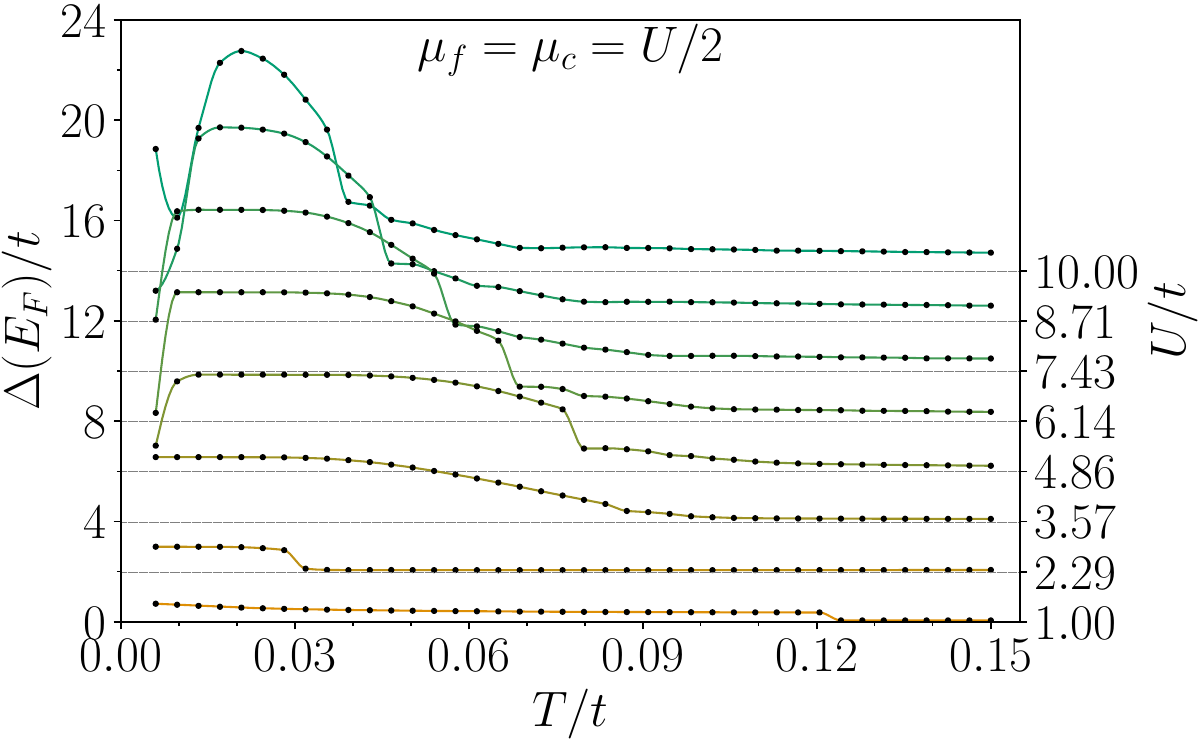}
    \end{subfigure}
    \hfill
    \begin{subfigure}{0.47\columnwidth}
        \includegraphics[width=\textwidth]{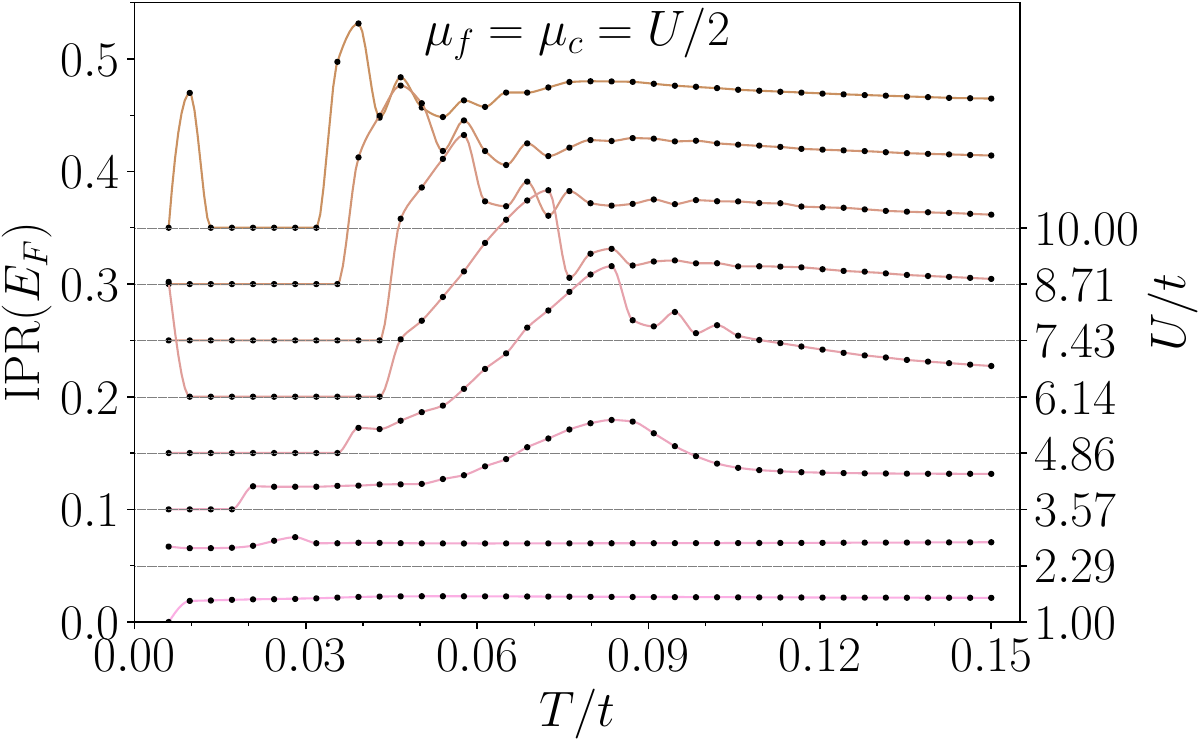}
    \end{subfigure}

    \caption{Specific heat $ c_v(T) $, c-DoS $ \rho_c(E_F, T) $, c-electron gap $ \Delta(E_F, T) $, and IPR($ E_F, T $) for the \mufix{} case on the triangular lattice. For clarity, curves corresponding to larger $U$ are vertically offset relative to the preceding one, where respective individual baselines are drawn for each curve.}
    \label{fig:N0_data}
\end{figure}

\begin{figure}[p!]
    \vspace{-1em}
    \centering
    \begin{subfigure}{0.47\columnwidth}
        \includegraphics[width=\textwidth]{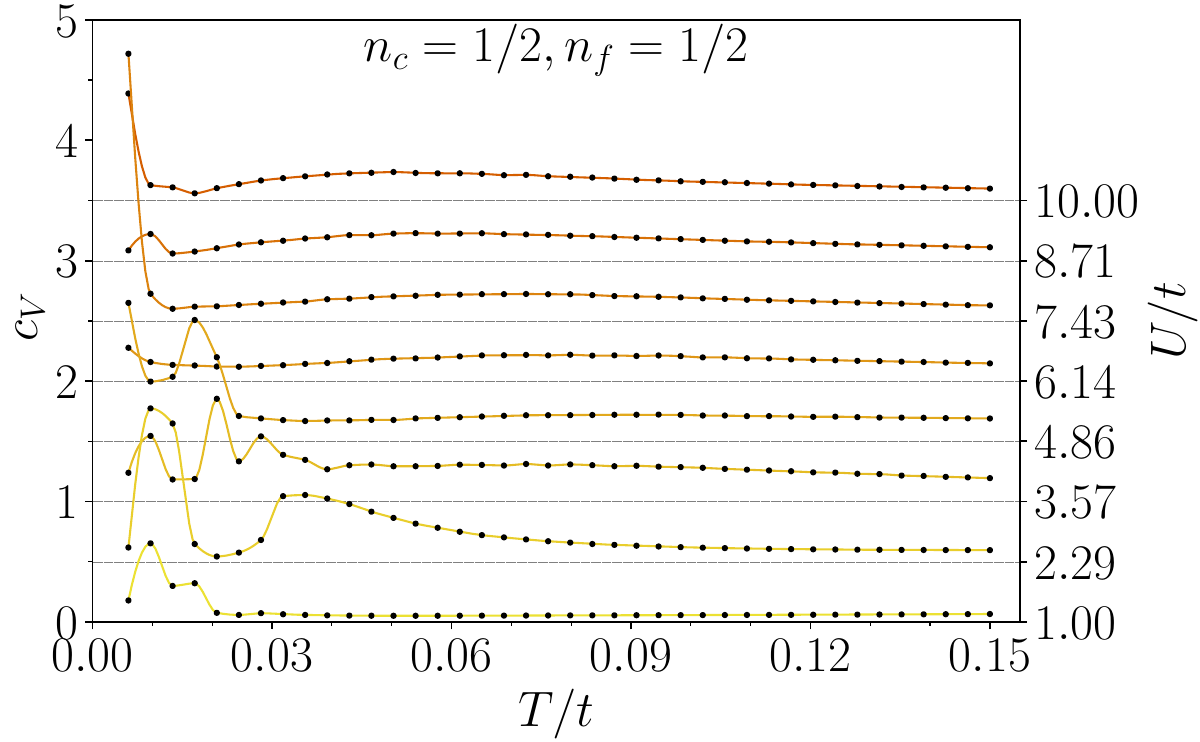}
    \end{subfigure}
    \hfill
    \begin{subfigure}{0.47\columnwidth}
        \includegraphics[width=\textwidth]{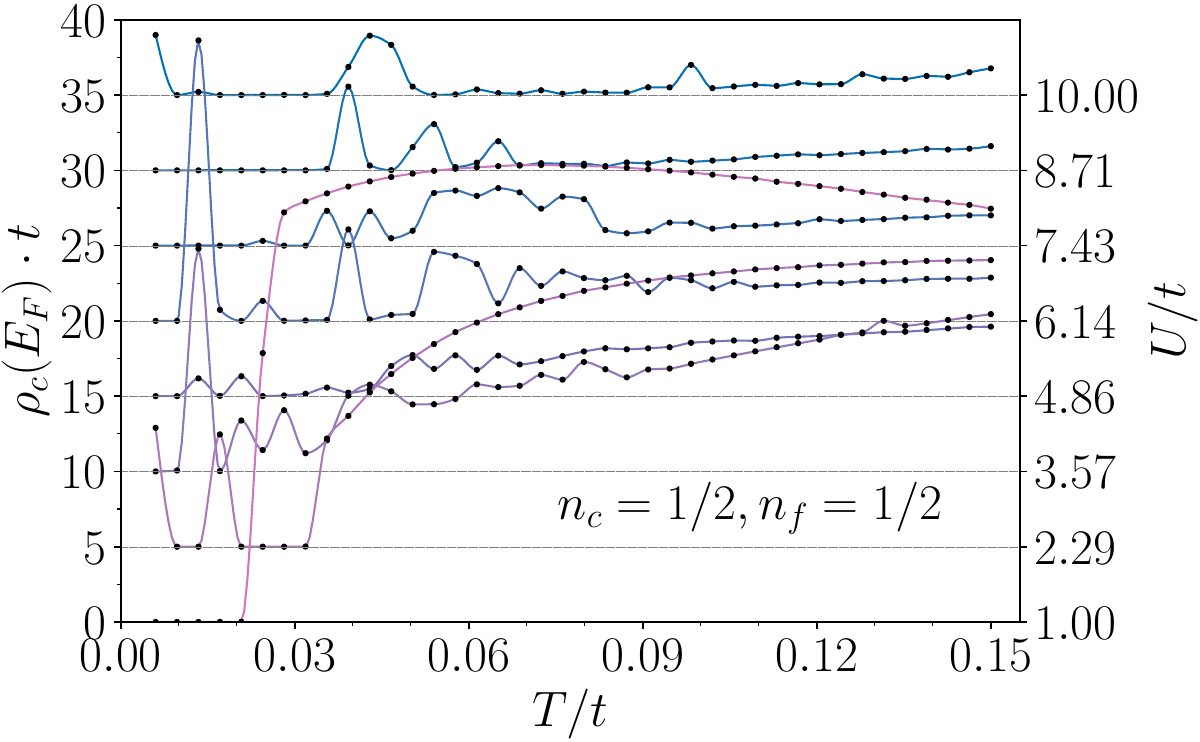}
    \end{subfigure}

    \begin{subfigure}{0.47\columnwidth}
        \includegraphics[width=\textwidth]{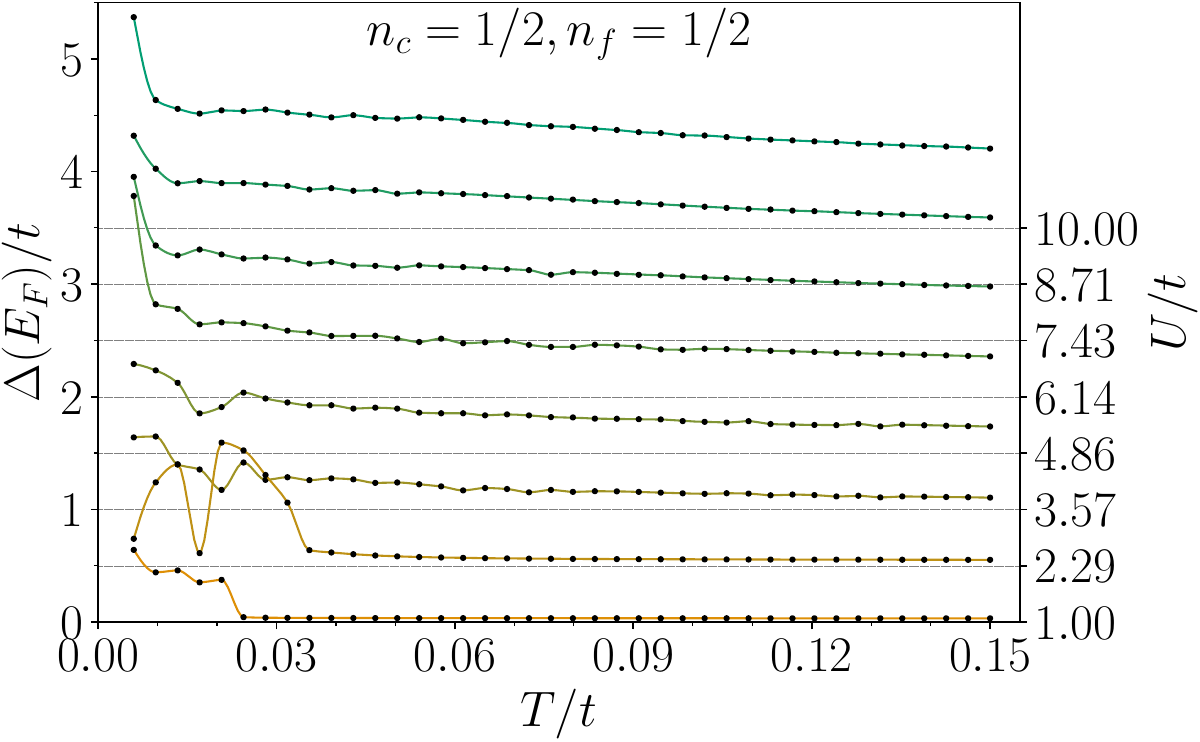}
    \end{subfigure}
    \hfill
    \begin{subfigure}{0.47\columnwidth}
        \includegraphics[width=\textwidth]{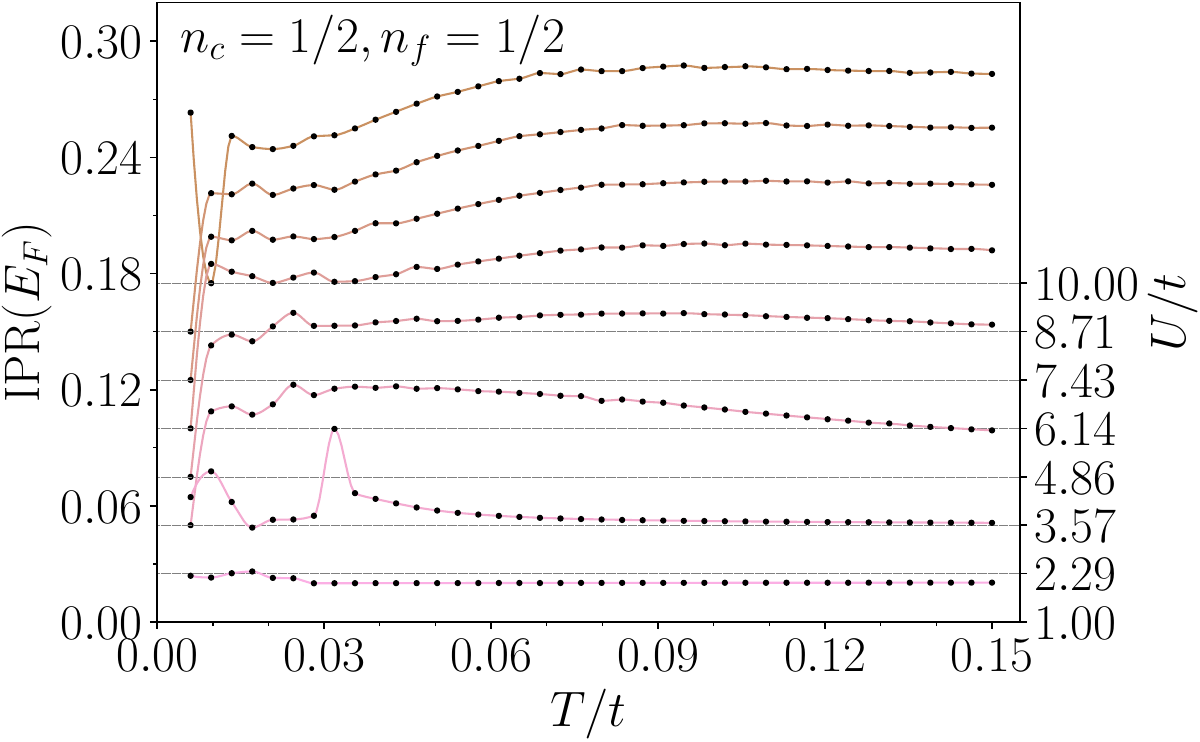}
    \end{subfigure}

    \caption{Specific heat $ c_v(T) $, c-DoS $ \rho_c(E_F, T) $, c-electron gap $ \Delta(E_F, T) $, and IPR($ E_F, T $) for the \hfcnv{} case on the triangular lattice. For clarity, curves corresponding to larger $U$ are vertically offset relative to the preceding one, where respective individual baselines are drawn for each curve.}
    \label{fig:N2_data}
\end{figure}

\begin{figure}[p!]
    \centering
    \begin{subfigure}{0.47\columnwidth}
        \includegraphics[width=\textwidth]{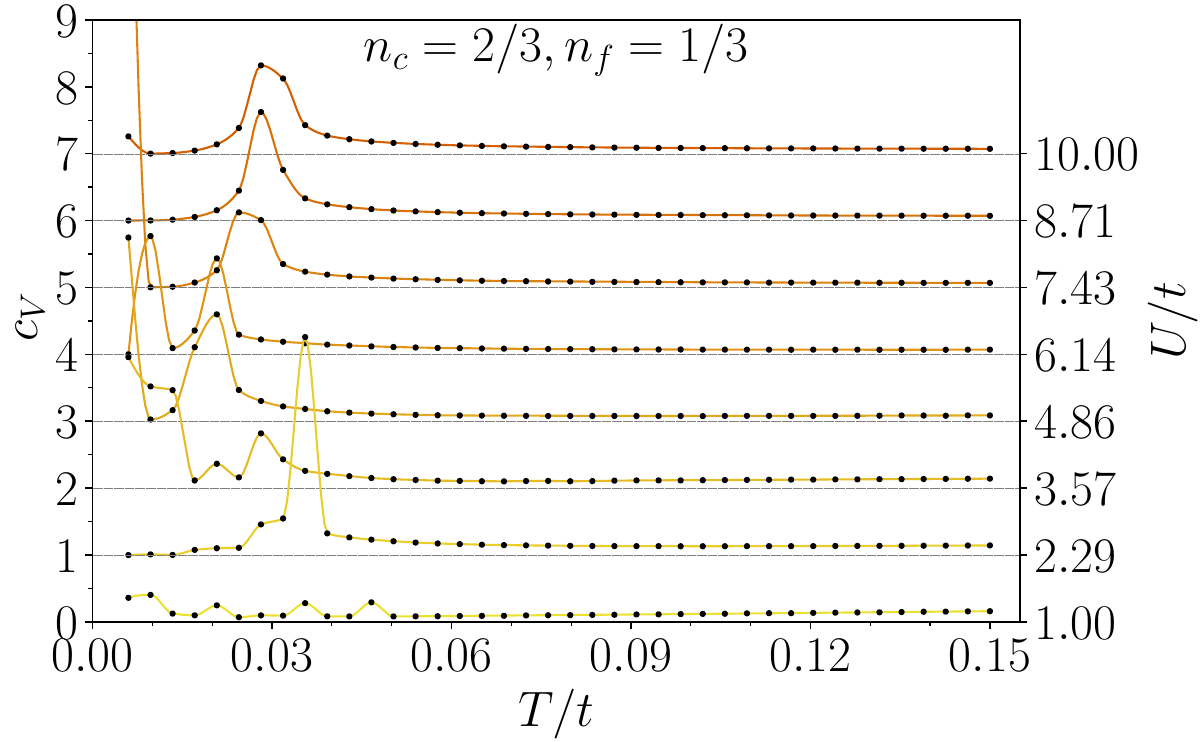}
    \end{subfigure}
    \hfill
    \begin{subfigure}{0.47\columnwidth}
        \includegraphics[width=\textwidth]{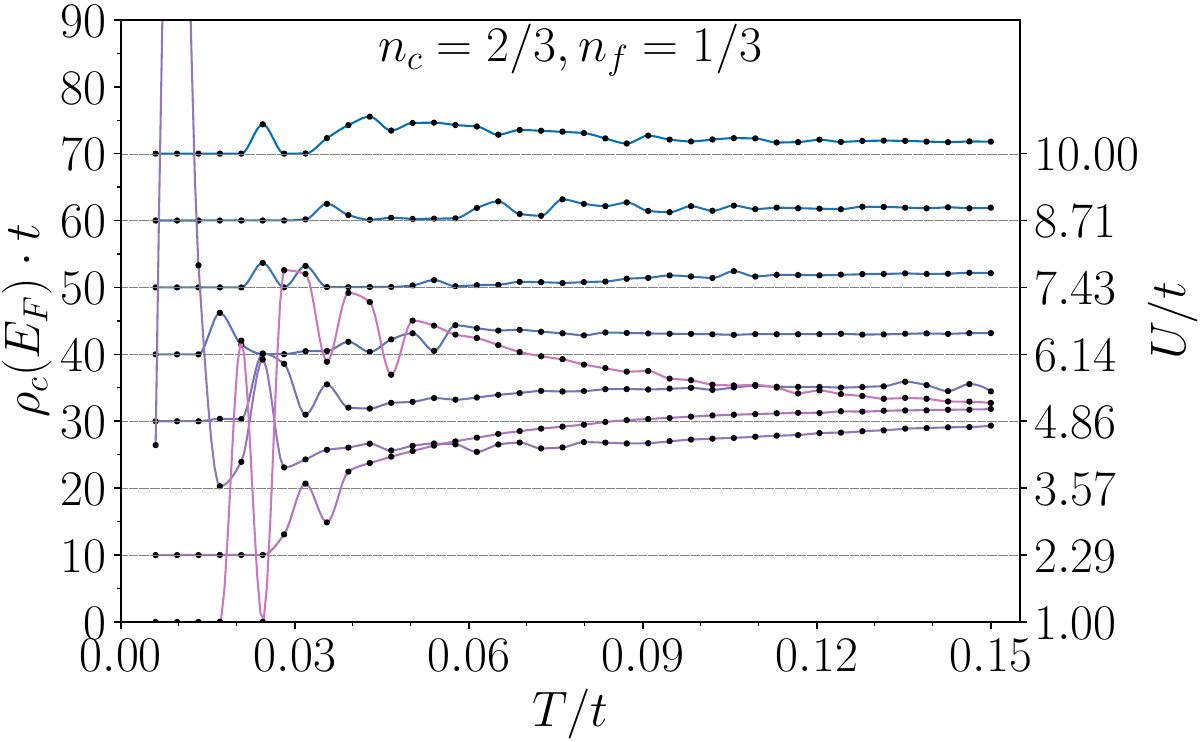}
    \end{subfigure}

    \begin{subfigure}{0.47\columnwidth}
        \includegraphics[width=\textwidth]{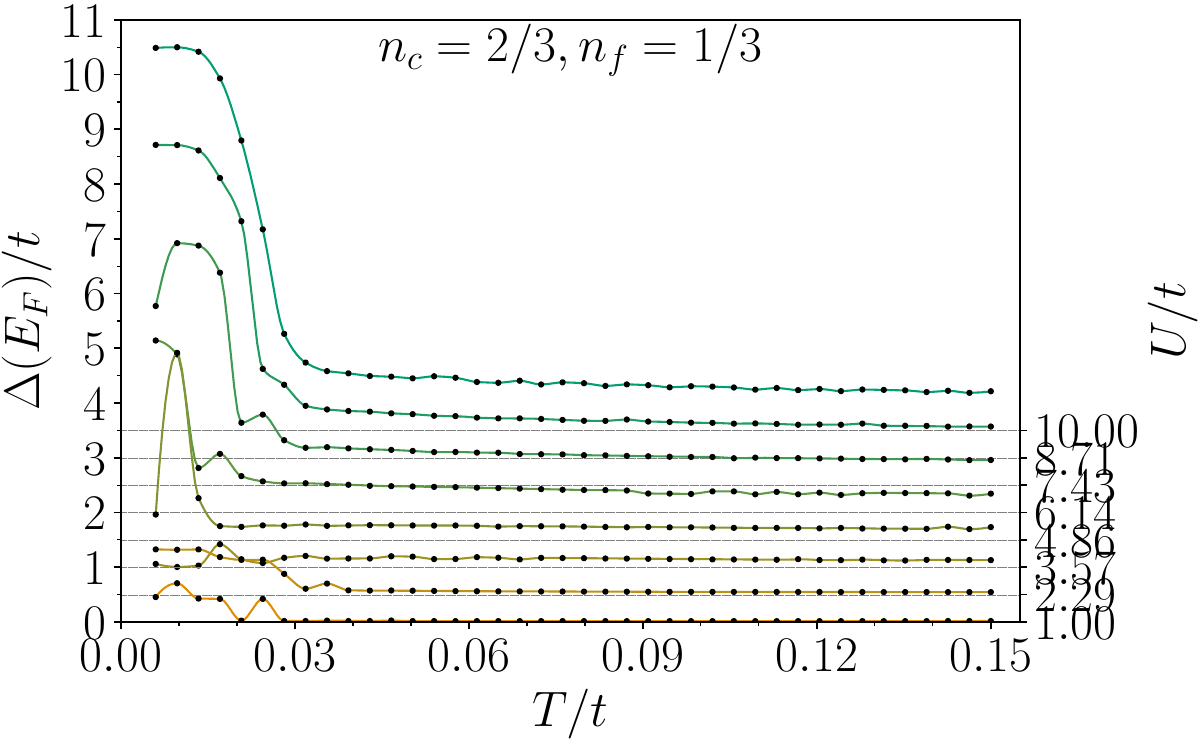}
    \end{subfigure}
    \hfill
    \begin{subfigure}{0.47\columnwidth}
        \includegraphics[width=\textwidth]{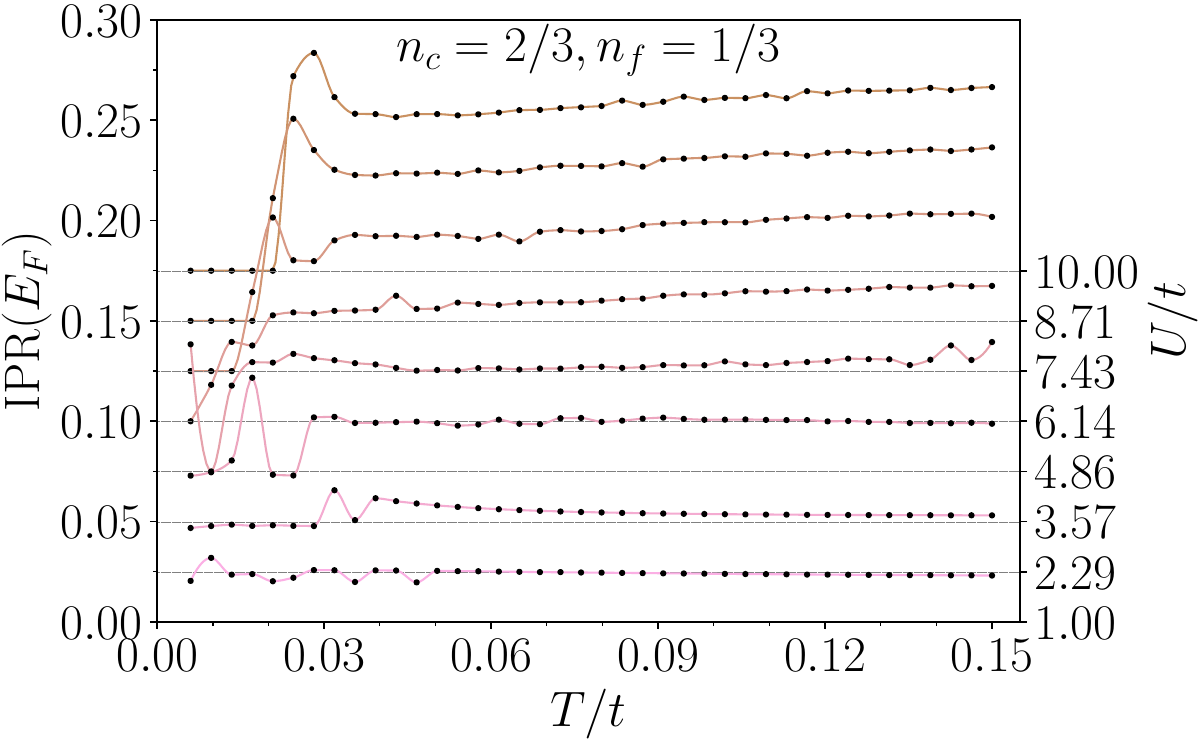}
    \end{subfigure}

    \caption{Specific heat $ c_v(T) $, c-DoS $ \rho_c(E_F, T) $, c-electron gap $ \Delta(E_F, T) $, and IPR($ E_F, T $) for the \hfgen{} case on the triangular lattice. For clarity, curves corresponding to larger $U$ are vertically offset relative to the preceding one, where respective individual baselines are drawn for each curve.}
    \label{fig:N1_data}
    \vspace{-1.1em}
\end{figure}

A detailed analysis of the observables for each case on the triangular lattice is presented in Appendix~\ref{app:triangular_tNNN0}; key findings are summarized below. The basic observables are shown in Figs.~\ref{fig:N0_data}--\ref{fig:N1_data}.

In all three cases, the specific heat exhibits peaks which signal the onset of charge ordering. At large interaction strengths, additional very-low-temperature features emerge in the specific heat, suggesting possible ground-state instabilities or quantum phase transitions at zero temperature, which lie beyond the applicability of the MC method. The Coulomb interaction progressively depletes the density of states at the Fermi edge and opens a gap, with a crossover from an Anderson-insulating to a Mott-insulating regime at intermediate interaction strengths, as evident from IPR($E_F$). The double-flavor occupation is strongly suppressed by the Coulomb interaction across the full temperature range of our simulations. For the \mufix{} case (Fig.~\ref{fig:N0_data}), at sufficiently high temperatures, the half-filling condition is automatically satisfied even without an a priori constraint on $ \bar{n}_f $ or $ \bar{n}_c $, and the triangular lattice increasingly resembles the square lattice in terms of the behavior of the occupation numbers.

\paragraph*{CDW ordering:} The three filling conditions lead to qualitatively different CDW behavior. The \mufix{} case (Fig.~\ref{fig:N0_data}) displays the most robust CDW phase: the specific heat peaks are broad, and the CDW exhibits universal scaling above a moderate interaction strength threshold. In the \hfgen{} case (Fig.~\ref{fig:N1_data}), fixing the particle densities sharpens the specific heat peak and shifts the onset of universality to a higher interaction strength threshold, while the Fermi-level gap is reduced compared to the \mufix{} case, reflecting a weaker nesting efficiency under the imposed filling constraint.
The \hfcnv{} (Fig.~\ref{fig:N2_data}) case exhibits the strongest suppression of charge ordering among the three cases: the sharp specific heat peak vanishes at large interaction strengths, replaced by a very broad feature, and the gap at the Fermi level is the smallest. This trend reflects the increasing frustration of charge order as the filling conditions depart from the grand-canonical condition toward conventional half-filling on the non-bipartite lattice.
As intuitively expected, increasing the ratio of localized f-electrons reduces the tendency of the system to form CDW which needs the delocalized c-electrons.

In all cases, for sufficiently large interaction strengths, the preferred ordering wavevector of CDW, $ \mathbf{q}_\text{max} $, coincides with the high-symmetry point $\mathbf{K} = \frac{1}{3}(\mathbf{b}_1 - \mathbf{b}_2)$ at the corner of the Brillouin zone of the triangular lattice. This reinforces the fact that the mechanism of CDW formation is driven by the Coulomb interaction (rather than phonons) through the common nesting mechanism~\cite{zhu_2017_charge_density_wave_origin}, which is consistent with the interpretation of the influence of next-nearest-neighbor hopping (see below).

Before the onset of the CDW transition, as $ U $ increases toward $ U_c $, a competition between different wavevectors is always observed; the susceptibility peaks initially appear at incommensurate wavevectors before locking onto the K-point at larger interaction strengths.\footnote{Ref.~\cite{freericks_exact_2003} shows that at small $U$ and away from half filling, incommensurate CDW phases exist whose ordering wavevector varies continuously with electron concentration.} This competition of wavevectors, without a single dominant one, is the hallmark of the frustrated charge-liquid crossover regime.
Moreover, the $\mathbf{q} = 0$ point (the $\Gamma$-point) is also among the wavevectors that maximize the charge susceptibility (though with a lower value than that of $\chi(\mathbf{q}_\text{max})$), indicating a possible charge instability and serving as a signature for an imminent Mott transition~\cite{kotliar_compressibility_2002}.

\subsubsection{Effect of finite next-nearest-neighbor hopping}
Introducing a perturbatively small next-nearest-neighbor hopping $ t_{\mathrm{NNN}} > 0 $ leaves the overall phase diagram intact but significantly affects the nesting quality and the universality of the CDW phase transition (see Appendix~\ref{app:triangular_NNN} for details). The phase borders are only slightly altered, confirming that the overall topology of the phase diagram, including the extent of the charge-liquid region, is a genuine feature of the model and not an artifact of perfect nesting.

In the \mufix{} case, the effect is most dramatic: the universal CDW regime is destroyed for all investigated interaction strengths once $t_\text{NNN}$ exceeds a small threshold, underscoring the sensitivity of the CDW scaling properties to Fermi-surface geometry in the grand-canonical ensemble.
In contrast, \hfgen{} and \hfcnv{} cases retain their qualitative phase structure, with the universality thresholds shifting but the CDW ordering and charge-liquid regions remaining intact.

We note that a finite $t_{\text{NNN}}$ warps the Fermi surface and thereby shifts the optimal nesting vector, which might be unachievable on a finite lattice. Part of the apparent loss of CDW universality at finite $t_{\text{NNN}}$ may therefore reflect this commensurability limitation rather than an intrinsic destruction of the ordering tendency. A definitive disentanglement of the two effects requires system sizes commensurate with the shifted nesting vector and is left for future work; the robustness of the phase borders reported above nonetheless indicates that the topology of the phase diagram is not affected by this issue.

\subsection{Kagome lattice}
%
\begin{figure}[tb]
    \centering
    \begin{subfigure}{0.47\columnwidth}
        \includegraphics[width=\textwidth]{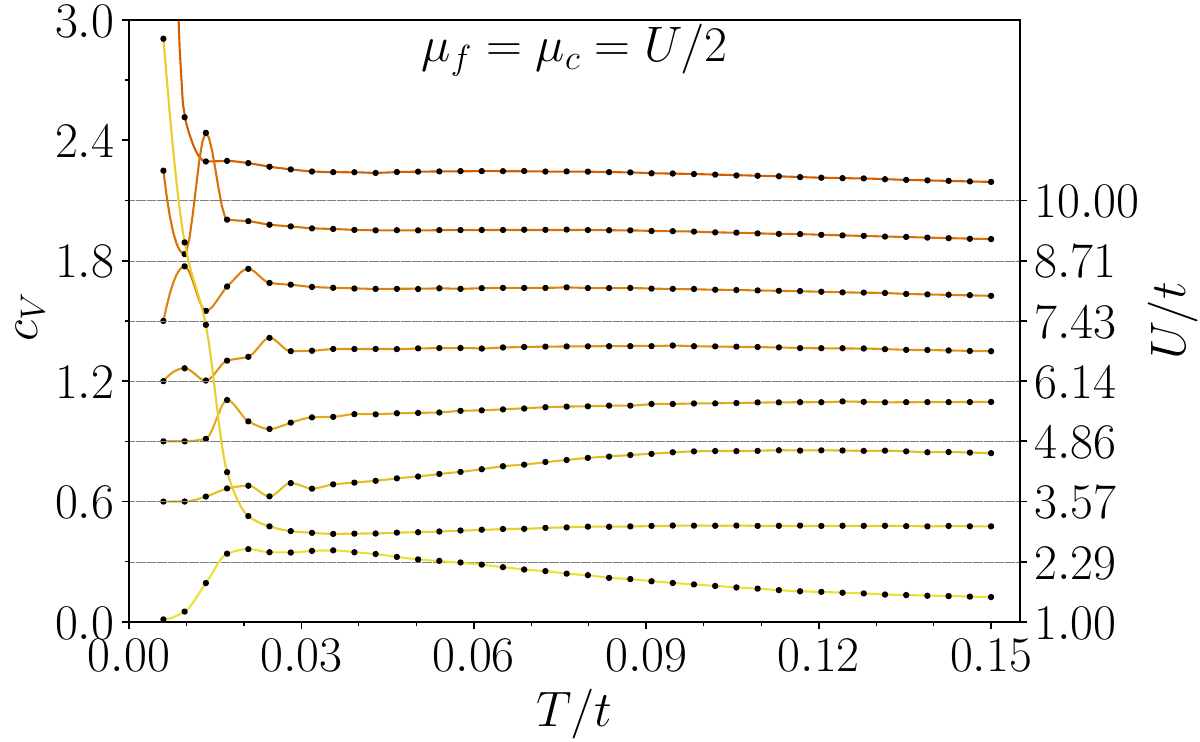}
    \end{subfigure}
    \hfill
    \begin{subfigure}{0.47\columnwidth}
        \includegraphics[width=\textwidth]{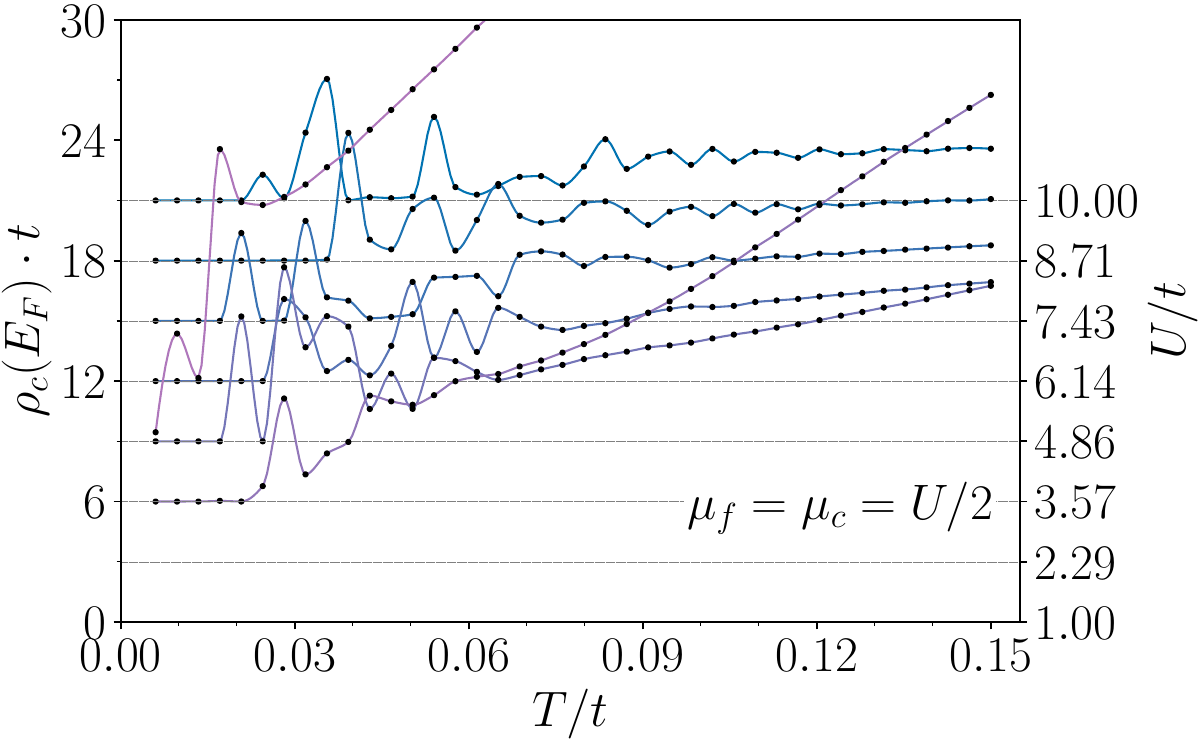}
    \end{subfigure}

    \begin{subfigure}{0.47\columnwidth}
        \includegraphics[width=\textwidth]{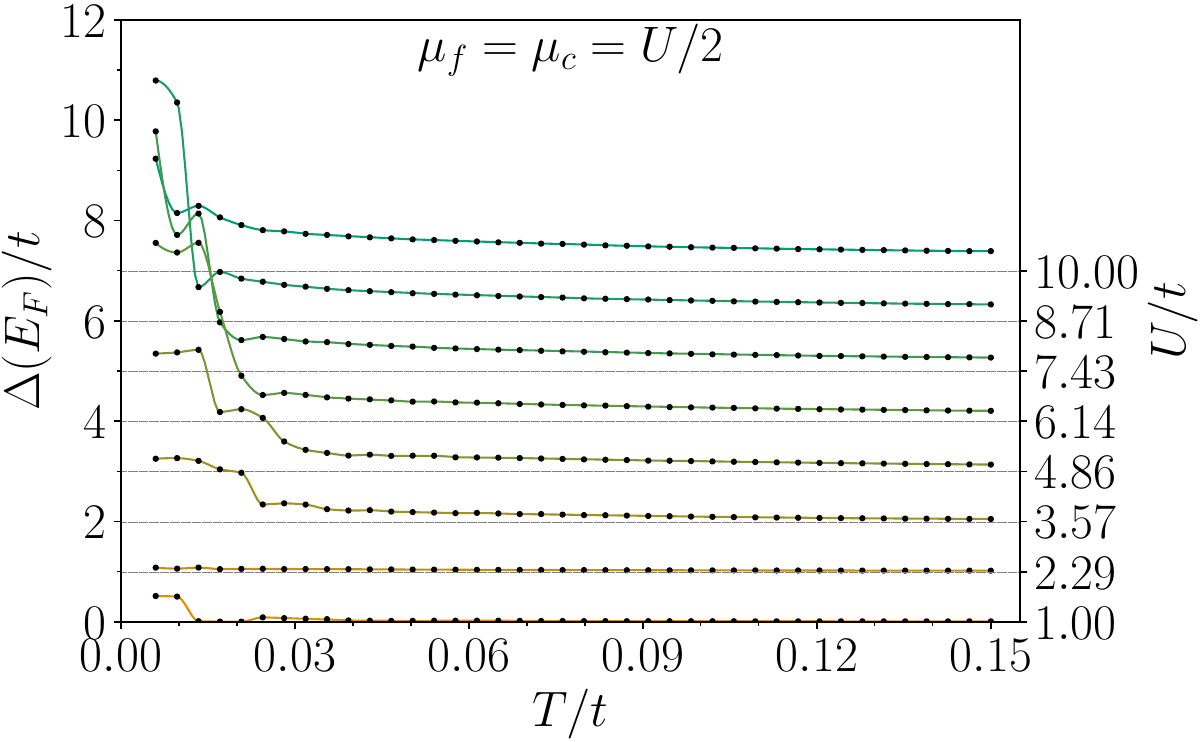}
    \end{subfigure}
    \hfill
    \begin{subfigure}{0.47\columnwidth}
        \includegraphics[width=\textwidth]{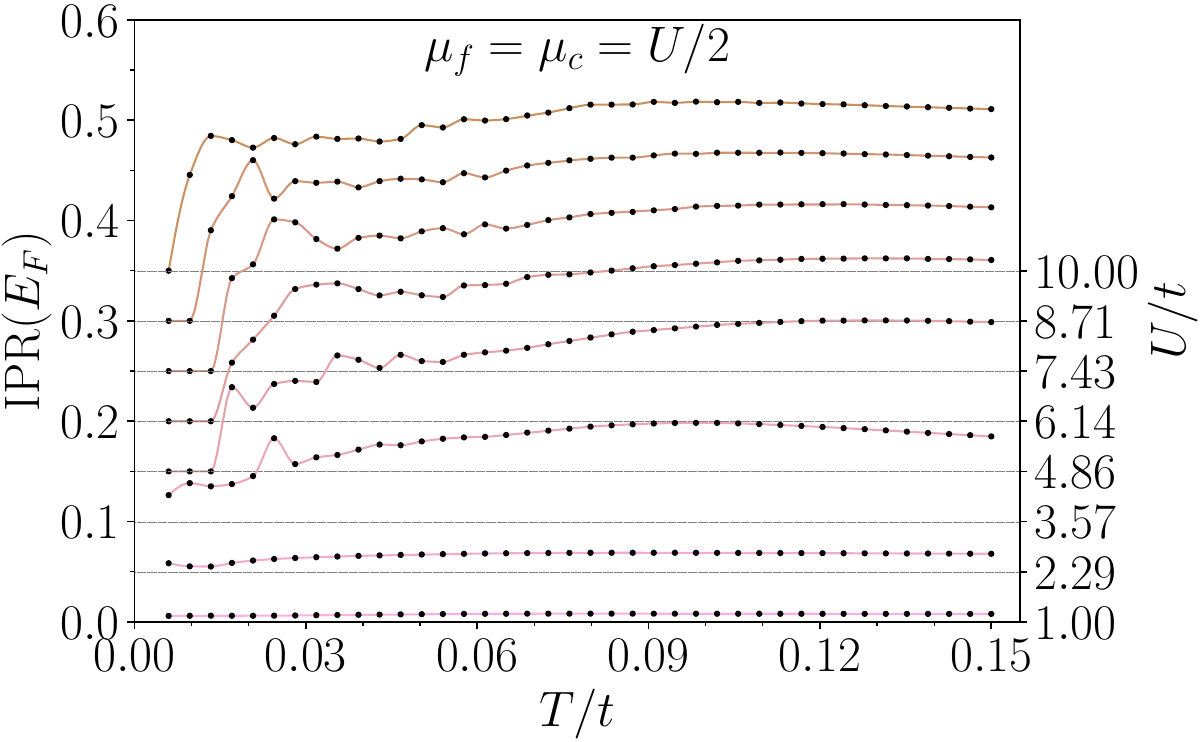}
    \end{subfigure}

    \caption{Specific heat $ c_v(T) $, c-DoS $ \rho_c(E_F, T) $, c-electron gap $ \Delta(E_F, T) $, and IPR($ E_F, T $) for the \mufix{} case on the kagome lattice. For clarity, curves corresponding to larger $U$ are vertically offset relative to the preceding one, where respective individual baselines are drawn for each curve.}
    \label{fig:N0_data_kagome}
\end{figure}
\begin{figure}[p!]
    \vspace{-1em}
    \centering
    \begin{subfigure}{0.47\columnwidth}
        \includegraphics[width=\textwidth]{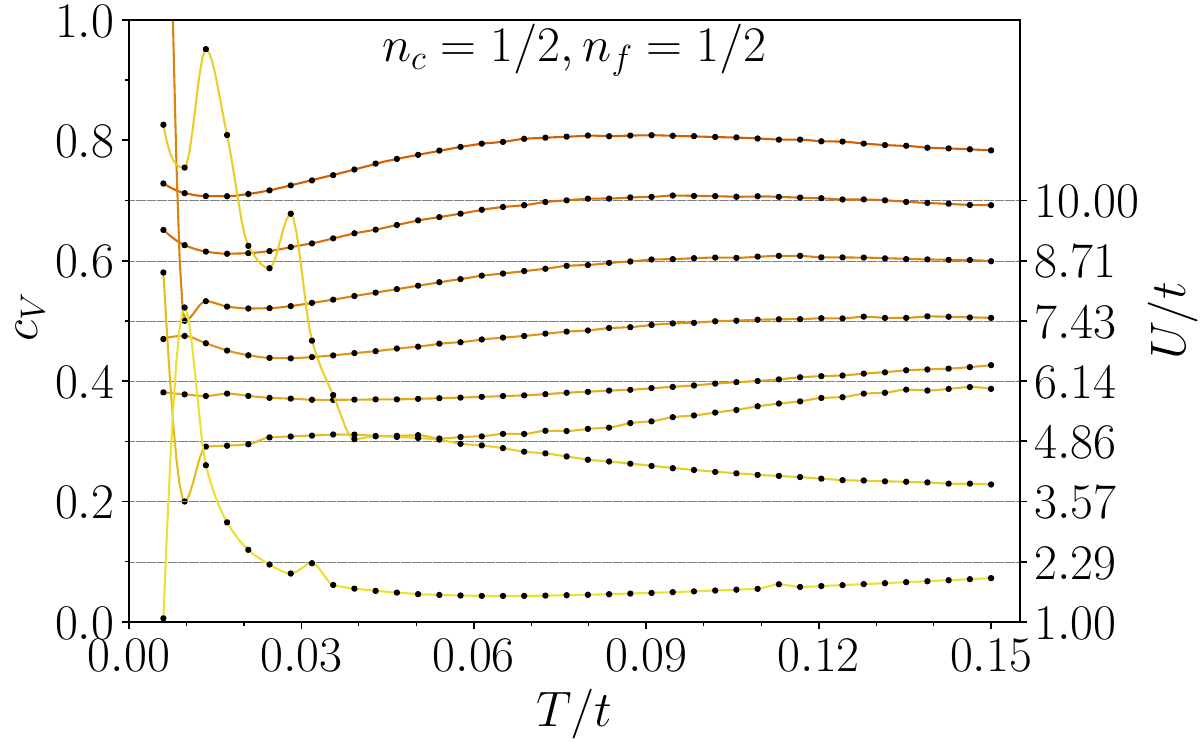}
    \end{subfigure}
    \hfill
    \begin{subfigure}{0.47\columnwidth}
        \includegraphics[width=\textwidth]{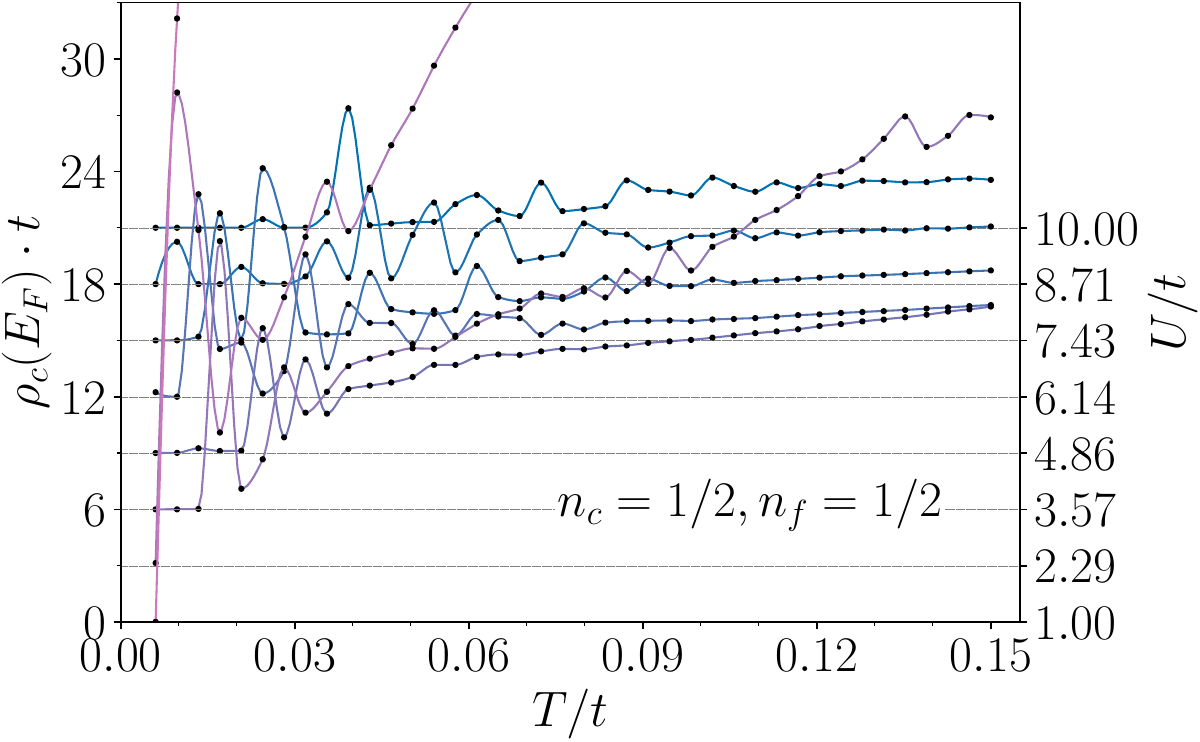}
    \end{subfigure}

    \begin{subfigure}{0.47\columnwidth}
        \includegraphics[width=\textwidth]{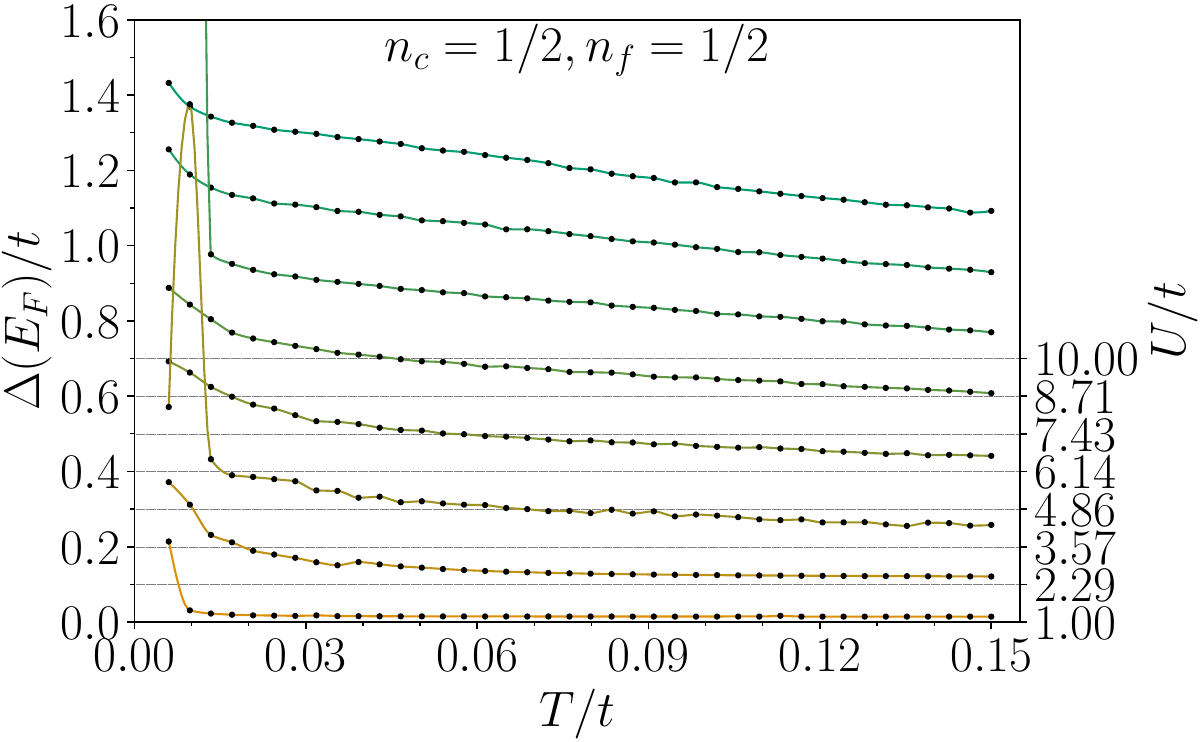}
    \end{subfigure}
    \hfill
    \begin{subfigure}{0.47\columnwidth}
        \includegraphics[width=\textwidth]{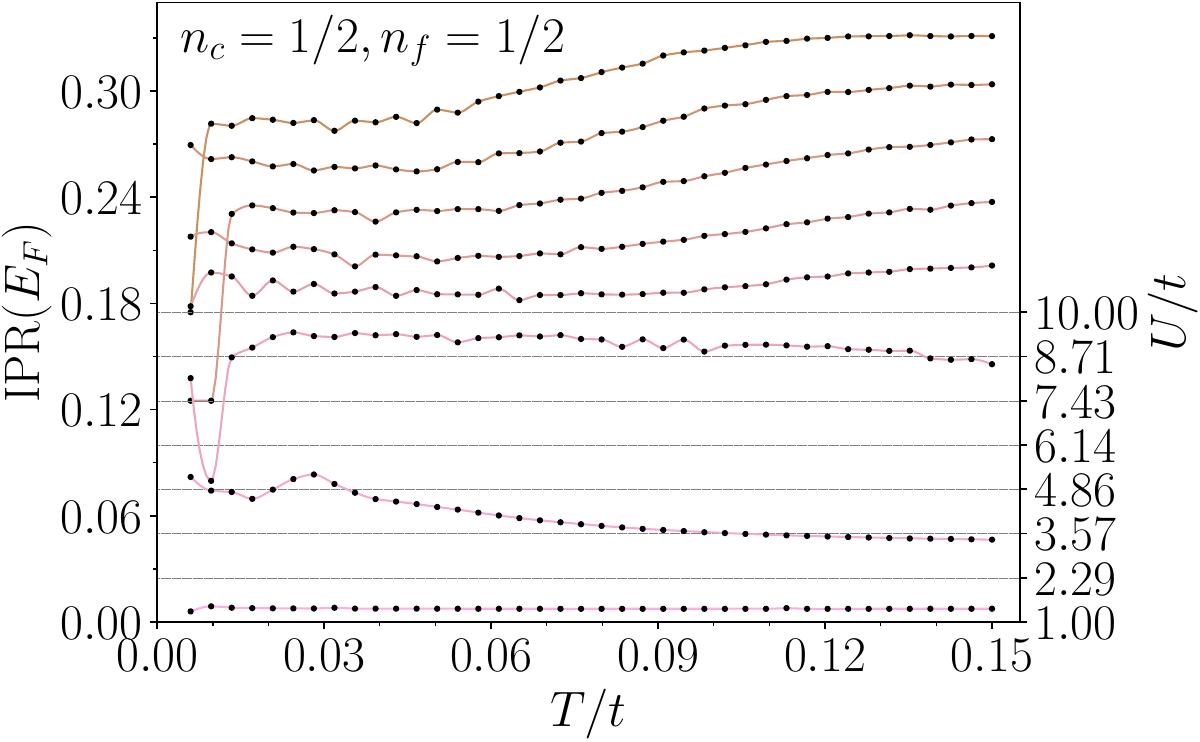}
    \end{subfigure}

    \caption{Specific heat $ c_v(T) $, c-DoS $ \rho_c(E_F, T) $, c-electron gap $ \Delta(E_F, T) $, and IPR($ E_F, T $) for the \hfcnv{} case on the kagome lattice. For clarity, curves corresponding to larger $U$ are vertically offset relative to the preceding one, where respective individual baselines are drawn for each curve.}
    \label{fig:N2_data_kagome}
\end{figure}
\begin{figure}[p!]
    \centering
    \begin{subfigure}{0.47\columnwidth}
        \includegraphics[width=\textwidth]{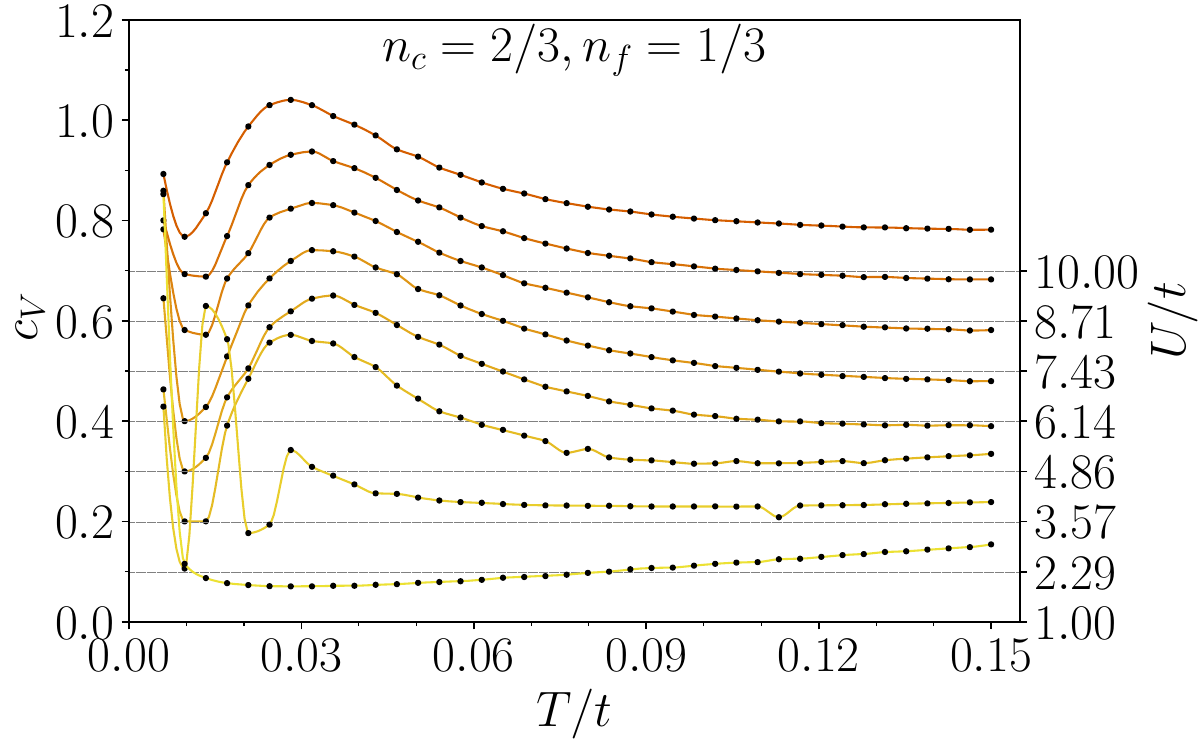}
    \end{subfigure}
    \hfill
    \begin{subfigure}{0.47\columnwidth}
        \includegraphics[width=\textwidth]{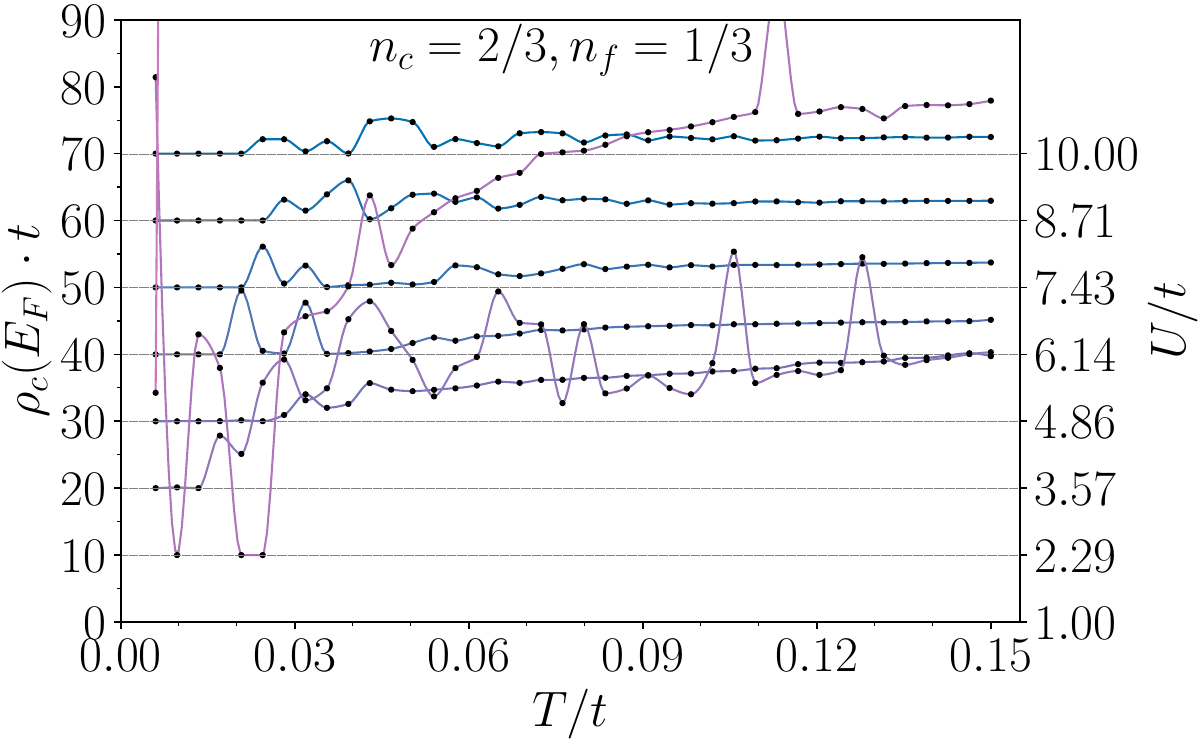}
    \end{subfigure}

    \begin{subfigure}{0.47\columnwidth}
        \includegraphics[width=\textwidth]{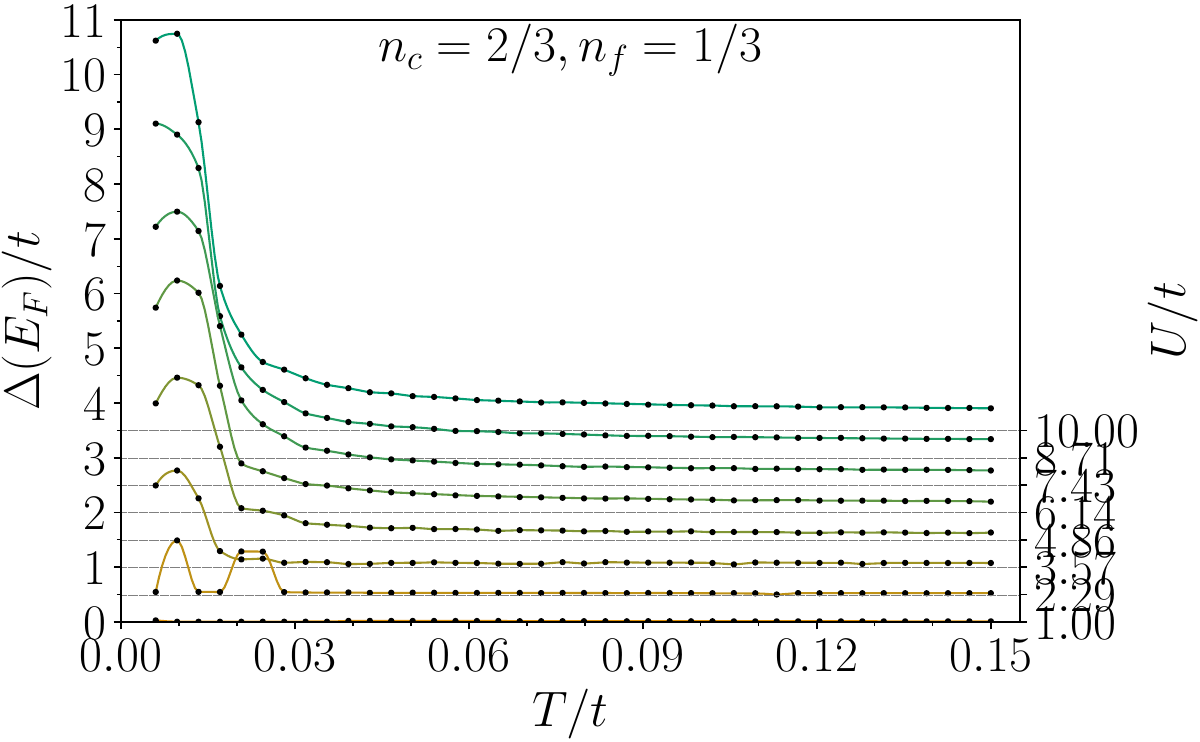}
    \end{subfigure}
    \hfill
    \begin{subfigure}{0.47\columnwidth}
        \includegraphics[width=\textwidth]{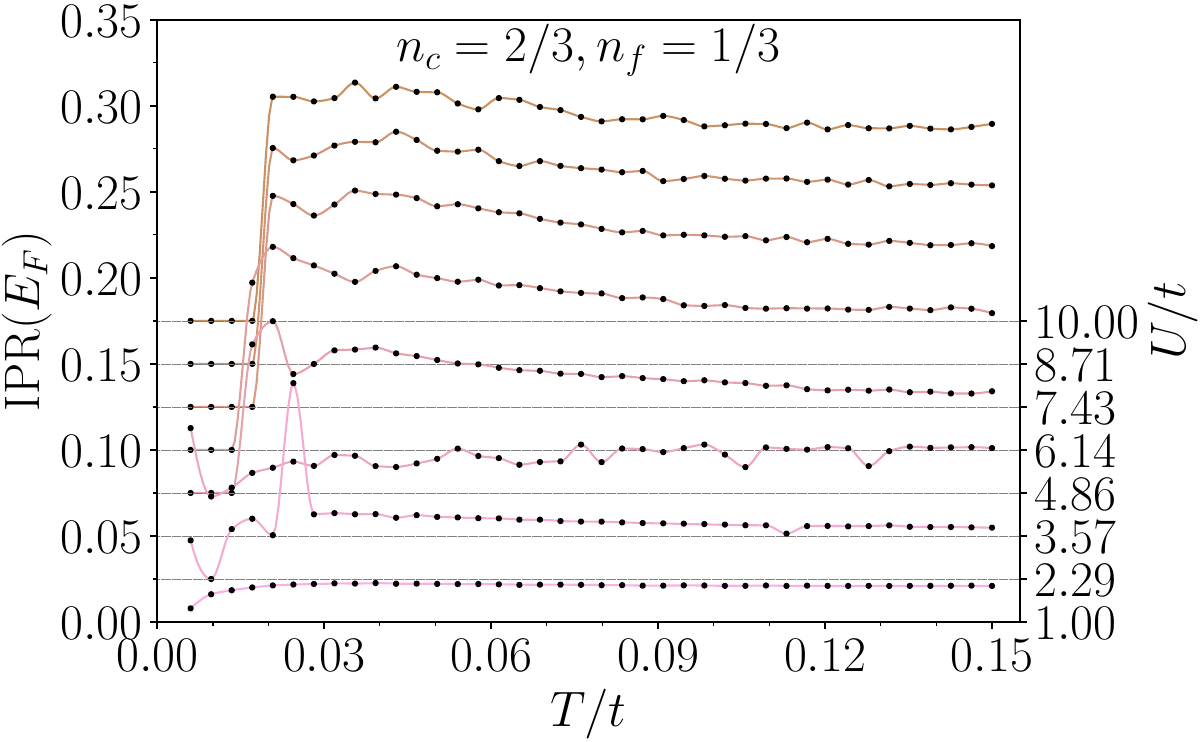}
    \end{subfigure}

    \caption{Specific heat $ c_v(T) $, c-DoS $ \rho_c(E_F, T) $, c-electron gap $ \Delta(E_F, T) $, and IPR($ E_F, T $) for the \hfgen{} case on the kagome lattice. For clarity, curves corresponding to larger $U$ are vertically offset relative to the preceding one, where respective individual baselines are drawn for each curve.}
    \label{fig:N1_data_kagome}
    \vspace{-1.1em}
\end{figure}
A detailed analysis of the observables for each filling condition is presented in Appendix~\ref{app:kagome}; key findings are summarized below. The basic observables are shown in Figs.~\ref{fig:N0_data_kagome}--\ref{fig:N1_data_kagome}.

The most striking observation on the kagome lattice is the complete absence of the CDW phase in all three cases.
In contrast to the triangular lattice, the specific heat peaks that signal CDW ordering are either absent or shifted to temperatures below the accessible range. In the \mufix{} case (Fig.~\ref{fig:N0_data_kagome}), no universal behavior is observed above the lowest accessible temperatures; at large interaction strengths, very-low-temperature features emerge that are consistent with a quantum phase transition at zero temperature. The \hfgen{} and \hfcnv{} cases (Fig.~\ref{fig:N2_data_kagome}--\ref{fig:N1_data_kagome}) show qualitatively similar behavior: the sharp specific heat peaks characteristic of the triangular lattice are replaced by broad, featureless maxima.

The IPR and density of states at the Fermi edge evolve qualitatively as on the triangular lattice, with localization increasing and states being depleted at the Fermi level as the Coulomb interaction grows, but the IPR values on the kagome lattice are somewhat lower, implying a less effective localization. The Fermi-level gap follows a pattern similar to the triangular lattice, with the gap being smallest in the \hfcnv{} case (Fig.~\ref{fig:N1_data_kagome}), though all kagome cases show gap values that remain below those of their triangular-lattice counterparts.

The charge susceptibilities show a stark increase across all wavevectors at the lowest accessible temperatures, with the maximum occurring at $\mathbf{q}_\text{max} = 0$ rather than at a finite wavevector. We emphasize that the quantity computed here is the \emph{classical} charge susceptibility (see Sec.~\ref{sec:method}), and it must not be identified with the physical charge susceptibility, which is governed by the full \emph{quantum} density-density response involving the density operators. Accordingly, the pronounced growth of the classical zero-mode susceptibility as $T \rightarrow 0$ signals an instability of the \emph{classical} total charge, which is fully consistent with the concurrent opening of a correlation gap at the Fermi edge. Taken together with the depletion of $\rho_c(E_F)$ and the finite Fermi-level gap, these observations are consistent with an insulating ground state, plausibly of the Mott type, in all three cases. A definitive identification would, however, require the (quantum) compressibility and the associated dynamical response. The apparent low-temperature divergences may additionally reflect a quantum phase transition at $T = 0$ with $\mathbf{q}_{\text{max}} = 0$ that is beyond the reach of the presently used method.

The absence of the CDW in the kagome lattice can be attributed to the inability to resolve the charge-order frustration due to the considerable alteration of the band structure and the Fermi surface after the addition of a basis. Further analysis, including other fillings and larger system sizes, is necessary to unveil the full physics of the FKM on the kagome lattice.

\section{Conclusion}

We have presented a systematic investigation of the stability of the phase diagram of the spinless Falicov-Kimball model on two-dimensional non-bipartite triangular and kagome lattices under varying conditions of chemical potential, filling, and lattice geometry. Three ensembles were studied: fixed chemical potentials (FCP) $\mu_f = \mu_c = U/2$, generalized half-filling (GHF) with $\bar{n}_f = 1/3$ and $\bar{n}_c = 2/3$, and conventional half-filling (CHF) with $\bar{n}_f = \bar{n}_c = 1/2$. For each case, on the triangular lattice, the effects of a perturbatively small next-nearest-neighbor hopping $t_{\text{NNN}}$ were examined. Furthermore, the triangular lattice was extended to the kagome lattice by adding a basis.

On the triangular lattice, all three cases host Anderson insulator (AI), Mott insulator (MI), and charge-density wave (CDW) phases in the low-temperature regime. The CDW phase persists to higher temperatures in \mufix{} case and is most suppressed in \hfcnv{} case, where the conventional half-filling on a non-bipartite lattice introduces the strongest frustration of charge ordering. In all cases, for a sufficiently large interaction strength ($U \gtrsim 6.1$), the preferred CDW ordering wavevector coincides with the high-symmetry point K at the corner of the Brillouin zone, confirming that the CDW is driven by Coulomb interactions through the nesting mechanism.

A key finding is the persistent appearance of a non-universal charge-liquid regime, which occupies the region of parameter space between the weakly correlated regime and the CDW phase, at moderate strengths of the Coulomb interaction, $ U < U_c $, though with varying extent. This regime is identified by the simultaneous absence of universal CDW scaling and the presence of broad, competing charge susceptibility peaks across multiple wavevectors, finite density of states at the Fermi edge, and partial localization of the c-electron wavefunctions. Its extent depends sensitively on the filling fractions and Fermi-surface geometry. We emphasize that we avoid distinguishing between a ``classical'' and ``quantum'' charge liquid as in Oliveira et al.~\cite{kirchner_2019_classical_quantum_liquid}.
The non-universal regime adjacent to the CDW phase is consistent with a \emph{quantum} charge liquid characterized by a multiply degenerate ground state or a dense low-energy manifold of quasi-degenerate excitations.

The introduction of a small next-nearest-neighbor hopping $t_{\text{NNN}}$ acts as a perturbation to the Fermi surface that does not drastically alter the overall phase boundaries but significantly affects the quality of nesting and, consequently, the universality of the CDW phase transition.
This sensitivity of the CDW universality to $t_{\text{NNN}}$ underscores the fragility of the nesting-driven CDW mechanism and is relevant to experimental realizations where perfect nearest-neighbor-only hopping is never achieved.

Transition from triangular to kagome lattice introduces a qualitative change: the most striking result is the complete absence of the CDW phase in all three cases. Instead, the system is an insulator plausibly of the Mott type, as suggested by the depleted density of states, the finite Fermi-level gap, and the growth of the \emph{classical} zero-mode ($\mathbf{q}_\mathrm{max} = 0$) susceptibility.\footnote{Ref.~\cite{freericks_exact_2003} identifies a continuous Mott-like metal-insulator transition at half filling, with the gap opening continuously at $U_c \sim 2t^*$ on the Bethe lattice ($U_c\sim 1.5t^*$ on the hypercubic lattice).}
The absence of CDW order can be attributed to a strong geometric frustration inherent in the kagome lattice, which prevents the formation of charge ordering. Notably, the localized and heavy flat c-band in the kagome tight-binding model enhances correlations but does not stabilize a CDW. Very-low-temperature divergences in both the susceptibility and specific heat suggest a possible quantum phase transition as $T \rightarrow 0$, which lies beyond the reach of the present Monte Carlo method and warrants investigation by complementary techniques.

We note several limitations of the present study that should be kept in mind when interpreting our results. The system sizes employed ($L \leq 12$, corresponding to $N_\mathrm{site} \leq 144$ for the triangular lattice and $N_\mathrm{site} \leq 432$ for the kagome lattice) are modest, constrained by the $\mathcal{O}(N_\mathrm{site}^3)$ diagonalization cost per MC step, the high cost for the imposition of the half-filling conditions and the sheer size of the scanned parameter space.
The identification of the charge-liquid regime rests on the observed absence of universal scaling behavior, which is inherently difficult to establish definitively from finite-size data since a finite-size crossover could, in principle, mimic a genuine non-universal regime; for this reason we deliberately characterize it as a frustrated finite-temperature crossover rather than as a distinct zero-temperature phase.
In both canonical cases (\hfgen{} and \hfcnv{}), fixing the occupations at very low temperatures becomes numerically difficult, especially for smallest $ U $-values.
At the lowest temperatures studied ($T \sim 0.01$), the acceptance rate for single-site f-electron updates drops steeply, and ergodicity of the Markov chain cannot be guaranteed without advanced sampling techniques such as parallel tempering or cluster updates. A rigorous determination of the integrated autocorrelation time at each point in the phase diagram (rather than the ad-hoc value of 20), along with error estimates from resampling, would strengthen the quantitative reliability of the reported phase boundaries. These considerations motivate the use of larger system sizes and enhanced sampling algorithms in future investigations. Comparison with semi-analytical methods like Hubbard approximations for the Mott transition, mean-field theories for the CDW transition, and strong-coupling $1 / U$ expansion~\cite{gruber_1992_largeU_expansion, gruber_macris_1996_falicov_kimball_exact} for the large-$U$ limit, is also required for corroborating the numerical results and a better physical understanding of the system.

Several avenues for future work emerge from this study. First, a systematic finite-size scaling analysis would enable the determination of the universality class of the CDW phase transition and the nature of the charge-liquid regime and its associated crossovers/transitions. A definitive characterization of the charge-liquid regime---including its spectral properties, the nature of its excitations, and its system-size scaling---demands further investigations. Second, the very-low-temperature regime ($T \ll 0.01$), where quantum phase transitions are anticipated, could be explored using semi-analytical methods. Third, the kagome lattice at alternative fillings and with finite $t_{\text{NNN}}$ remains largely unexplored and may host novel phases, including (exotic) frustrated charge-liquid regimes.

Finally, as venues for experimental realization, we note that the FKM could serve as a simple model for two-dimensional layered transition metal dichalcogenide compounds~\cite{qiao_2017_mottness_cdw_tmdc, manzeli_2017_2d_transition_metal_dichalcogenides, hwang_2024_charge_density_wave_2d_transition_metal_dichalcogenides, kumar_2023_mott_insulator_tmdc}, to describe their Mott and CDW phases, as well as a model for conceiving and developing quantum devices based on the manipulation of charge-density waves~\cite{balandin_2021_charge_density_wave_quantum_device}.

\subsection*{Acknowledgements}

We thank Regine Frank and Sajjad Azizi for stimulating discussions during the initial phase of this work. ML gratefully acknowledges the hospitality of the University of Bonn, where part of this work was initiated.

\clearpage
\begin{appendices}

\section{Temperature-dependent smoothing}
\label{app:smoothing}

To obtain c-DoS and IPR as continuous functions of energy, $ \omega $,
a simple temperature-dependent smoothing weight
$ W_T(\varepsilon) = 1 - \text{tanh}^2( \frac{\varepsilon}{2 T} )
= \text{sech}^2(\frac{\varepsilon}{2 T}) $ is used, which has a peak at $ \varepsilon = 0 $ with a width roughly equal to $ T $.
The physical motivation for this is that the same factor appears in the expression for $ c_v(T) $ and also in the equation for the dI/dV signal of the spectroscopic scanning-tunneling microscopy in many-body quantum theory. A smoothed function is obtained via

\begin{equation}
  \tilde{f}(\omega) = \frac{\sum_{l \in \text{lattice}} W_T(\varepsilon_l - \omega) \, f(\varepsilon_l)}
  {\sum_{i \in \text{lattice}} W_T(\varepsilon_i - \omega)},
\end{equation}
where $ \varepsilon_{l, i} $ are the eigenenergies of $ \mathcal{H}_c $.
The smoothed $\rho_c(\omega)$ is always normalized so that
$ \int \mathrm{d} \omega \, \rho_c(\omega) = 1 $.

\section{Parameter space}
\label{app:parameter_space}

The system sizes considered in this work are $L = 8, 10, 12$. The temperature range is $ T \in [0.01, 0.15] $ and that of the Coulomb interaction strength is $ U \in [1.0, 10.0] $.
Below $ T = 0.01 $, the required number of MC steps increases substantially, and semi-analytical approximations might provide better results. The lower end of this range is also set by the finite-size single-particle level spacing: as discussed in Sec.~\ref{sec:method}, the temperatures analyzed are comparable to or larger than the interaction-renormalized level spacing throughout the bulk of the range, so that thermal broadening remains at least of the order of the discretization scale, whereas the very lowest temperature $T \sim 0.01\,t$ marks the onset of the regime where discretization effects become competitive.
The interaction strengths below $ U < 1.0 $ are in a perturbative range of
the tight-binding limit (the limit of a degenerate Fermi gas) or a weakly correlated regime\footnote{
Drawing on the scaling theory of localization~\cite{abrahams_1979_scaling_theory_localization},
if the spinless FKM on a non-bipartite lattice is tentatively placed in the ``orthogonal symmetry class'' (2d systems with time-reversal symmetry and no spin-orbit coupling), then the scaling theory predicts, for all disorder strengths, that \emph{all} states are localized --- no true metallic phase exists. However, this argument is not rigorously valid as the scaling theory is formulated strictly for \emph{quenched} disorder, not the effective annealed disorder in the FKM case.},
and $ U $-values above 10.0 are uninteresting as they are not physically realisable in conventional experiments, although they can be used for testing the MC method in the large-$ U $ limit. The strength of the nearest-neighbor hopping, $ t $, is used as a scale for other parameters such that the simulations internally utilize only dimensionless variables. In those units, the hopping strength is always $ t/t = 1$.
The next-nearest-neighbor hopping varies in the range $t_{\text{NNN}} / t \in [0, 0.1]$.

Imposing periodic boundary conditions leads to a discrete set of allowed wavevectors,
$ \mathbf{q}_{[h_1, h_2]} = \frac{1}{L} \left ( h_1 \mathbf{b}_1 + h_2 \mathbf{b}_2 \right) $, where $ h_{1,2} $ is an integer in $ [0, L) $.
The number of wavevectors increases with the linear size of the system as $ L^2 $. The susceptibilities and Binder cumulants must be computed for all these wavevectors.

\section{Detailed analysis: Triangular lattice, \texorpdfstring{$t_\text{NNN} = 0$}{vanishing next-nearest-neighbor hopping}}
\label{app:triangular_tNNN0}

The basic observables, specific heat $ c_v(T) $, density of states at the Fermi edge $ \rho_c(E_F, T) $, gap at the Fermi edge $ \Delta(E_F, T) $, and inverse participation ratio at the Fermi edge IPR($E_F, T$), for the three filling cases on the triangular lattice with vanishing next-nearest-neighbor hopping are shown in Figs.~\ref{fig:N0_data}--\ref{fig:N1_data}. We provide a detailed discussion of each case below.

\subsection{Case \texorpdfstring{\mufix{}}{FCP}: Fixed chemical potentials, \texorpdfstring{$\mu_f = \mu_c = U/2$}{both equal to U/2}}
Relatively broad peaks appear in the specific heat, whose positions vary with the system size $L$ and shift to lower temperatures as $U$ increases, similar to the behavior observed for the square lattice. For $U \gtrsim 4.8$, an additional very-low-temperature peak emerges, likely signaling an instability of the ground state (possibly a quantum phase transition) at zero temperature.
The localization measure IPR$(E_F)$ vanishes at the lowest temperatures due to the gap at the Fermi edge, rises to a peak around $T \sim 0.04$, and then stabilizes to a high-temperature value that increases with $U$ (from 0.03 at $U = 1.0$ to 0.12 at $U = 10.0$).\footnote{The high-temperature values are obtained by averaging
over the values at $ T \sim 0.15 $ (upper limit of the temperature range) for different lattice sizes.}

The c-electron density of states at the Fermi edge, $\rho_c(E_F)$, vanishes at very low temperatures ($T \lesssim 0.01$) due to either a CDW or Mott gap, and shows fluctuations across the transition due to the emergence of very localized subgap states. Beyond the transition, localized Hubbard bands form around the gap. The value of $ \rho_c(E_F) $ then stabilizes for higher temperatures $ T \gtrsim 0.08 $.
This ``high-temperature'' value decreases by increasing the Coulomb interaction strength, as the Coulomb interaction depletes the states at the Fermi edge (from 9.2 at $ U = 1.0 $ to 0.75 at $ U = 10.0 $).

For larger Coulomb interaction strengths, $ U \gtrsim 6.1 $, a gap, $ \Delta_\mufix $, develops in c-DoS at the Fermi edge, beginning at a value roughly equal to the Coulomb interaction strength and reaching a value roughly equal to the nearest-neighbor hopping strength, $ t = 1 $, at higher temperatures.
This ``asymptotic'' value increases with increasing Coulomb interaction
(from 0.5 at $ U = 6.1 $ to 0.9 at $ U = 10.0 $).

The average occupations of f- and c-electrons, $ \bar{n}_f $ and $ \bar{n}_c $, tend to their fixed values at higher temperatures, with the total occupation, $ \bar{n}_f + \bar{n}_c $, approaching 1; therefore, at the higher temperatures,
the half-filling condition is \emph{automatically} satisfied even without a priori fixing $ \bar{n}_f $ or $ \bar{n}_c $.
The double-flavor occupation is strongly suppressed by the Coulomb interaction in the studied temperature range.
Essentially, as the temperature and interaction strength are increased, the triangular lattice behaves increasingly similar to the square lattice, in terms of occupations.

Charge susceptibilities, $ \chi_{c,f}(\mathbf{q}, T) $, show peaks consistently at
$ \mathbf{q}_{\text{max}} \approx \frac{1}{3} (\mathbf{b}_1 \pm \mathbf{b}_2) $ for $ U \geq 2.3 $.
\footnote{Note that for $ L \neq 12 $, the \emph{exact} condition
$ \mathbf{q}_{\text{max}} = \frac{1}{3} (\mathbf{b}_1 \pm \mathbf{b}_2) $
cannot be satisfied, due to restricted allowed values of wavevectors, which is a consequence of the imposed periodic boundary conditions. For the CDW ordering wavevector $\mathbf{K}$, only system sizes with $L$ divisible by 3 are commensurate with the CDW supercell, which limits the number of strictly valid data points for finite-size analysis in the ordered region.}
The temperature for maximum susceptibility, $ T_{\chi_\text{max}} $, begins at 0.09 for $ U = 2.3 $ and reduces to 0.06 as $ U $ increases.
Universality is observed for $ U \geq 2.3 $.

\subsection{Case \texorpdfstring{\hfgen{}}{GHF}: Generalized half-filling, \texorpdfstring{$\bar{n}_f = 1/3$, $\bar{n}_c = 2/3$}{nf = 1/3, nc = 2/3}}
Fixing the average occupations changes the behavior qualitatively. The specific heat develops a relatively sharp peak in the range $T \in [0.025, 0.03]$. For $U \gtrsim 3.5$, an additional very-low-temperature peak appears.
The IPR$(E_F)$ behavior is similar to the \mufix{} case, though its peak shifts to a lower temperature, $T \sim 0.03$. The asymptotic IPR values increase steadily by increasing the Coulomb interaction strength (from 0.03 at $ U = 1.0 $ to 0.1 at $ U = 10.0 $).

Similar to the \mufix{} case, $ \rho_c(E_F) $ vanishes at very low temperatures, $ T \lesssim 0.01 $ and $ U \gtrsim 4.8 $, and shows fluctuations as the temperature approaches a transition; the nature of the emerging subgap states is localized, indicated by their relatively high IPR measures.
The value of DoS($ E_F $) then stabilizes for higher temperatures, $ T \gtrsim 0.06 $.
This high-temperature value decreases by increasing the Coulomb interaction strength,
as the Coulomb interaction depletes the states at the Fermi level
(from 26.0 at $ U = 1.0 $ to 1.3 at $ U = 10.0 $).

The dimensionless gap at the Fermi level, $\Delta_\hfgen / t$, is smaller than $\Delta_\mufix / t$ and remains below unity at higher temperatures. This asymptotic value increases with increasing Coulomb interaction
(from 0.3 at $ U = 4.8 $ to 0.9 at $ U = 10.0 $).

Charge susceptibilities show peaks at $\mathbf{q}_\mathrm{max} \approx \frac{1}{2}(\mathbf{b}_1 \pm \mathbf{b}_2)$ for $U \approx 4.8$, shifting to $\frac{1}{3}(\mathbf{b}_1 \pm \mathbf{b}_2)$ for $U \gtrsim 6.1$, with universality observed for $U \gtrsim 4.8$.

Fixing the occupations at very low temperatures ($T \lesssim 0.01$) becomes numerically difficult, and imposing the exact condition becomes infeasible near the tight-binding limit ($U \rightarrow 0$).

\subsection{Case \texorpdfstring{\hfcnv{}}{CHF}: Conventional half-filling, \texorpdfstring{$\bar{n}_f = \bar{n}_c = 1/2$}{nf = nc = 1/2}}
The specific heat behavior differs qualitatively from the \mufix{} and \hfgen{} cases: a relatively sharp peak appears near $T \sim 0.02$ for $U \lesssim 4.8$, which vanishes above $U \gtrsim 6.1$, leaving only a very broad peak. A sharp, very-low-temperature peak appears for $U \gtrsim 6.1$, indicating a possible zero-temperature instability.
The IPR$(E_F)$ also differs: the low-temperature peak ($T \sim 0.01$) has a lower value than the high-temperature limit. Similar to the \mufix{} and \hfgen{} cases, the asymptotic IPR values increase steadily by increasing the Coulomb interaction strength (from 0.03 at $ U = 1.0 $ to 0.1 at $ U = 10.0 $).

Similar to the \hfgen{} case, $ \rho_c(E_F) $ is vanishing at very low temperatures, $ T \lesssim 0.01 $,
shows fluctuations as the temperature approaches a transition and then stabilizes
for higher temperatures $ T \gtrsim 0.06 $.
This high-temperature value decreases by increasing the Coulomb interaction strength,
as the Coulomb interaction depletes the states at the Fermi level
(from 18.0 at $ U = 1.0 $ to 0.8 at $ U = 10.0 $).

The gap at the Fermi level, $\Delta_\hfcnv < \Delta_\hfgen$, remains small, indicating a suppression of the CDW phase in the studied temperature range.

Initially, for $U < 6.1$, many different wavevectors compete to maximize the susceptibility (e.g., $\mathbf{q}_\mathrm{max} \approx \frac{1}{4}(\mathbf{b}_1 \pm \mathbf{b}_2)$), while for $U \gtrsim 6.1$, the susceptibility peaks consolidate at $\frac{1}{3}(\mathbf{b}_1 \pm \mathbf{b}_2)$, with universality observed above this threshold.

Fixing the occupations at very low temperatures $ T \lesssim 0.01 $ becomes difficult numerically, and imposing the condition becomes impossible, probably due to the inherent nature of the ground-state close to the tight-binding limit, $ U \rightarrow 0 $.

\section{Detailed analysis: Triangular lattice, \texorpdfstring{$t_\text{NNN} > 0$}{finite next-nearest-neighbor hopping}}
\label{app:triangular_NNN}

In all cases, the phase borders are only slightly altered by a finite next-nearest-neighbor hopping $t_{\text{NNN}}$. The hopping $t_{\text{NNN}}$ acts as a small perturbation to the Fermi surface, which does not lead to drastic changes in the phase diagram, especially in the strong-coupling regime. It does, however, influence the quality of nesting, thereby affecting the scaling properties within the CDW phase.

In all three cases, the CDW phase exhibits universality only above a certain interaction strength threshold $U_c$. In the absence of next-nearest-neighbor hopping, $ t_{\mathrm{NNN}} = 0 $, these thresholds are $U^\mufix_c \sim 2.3$, $U^\hfcnv_c \sim 6.1$, and $U^\hfgen_c \sim 4.8$.
With finite $t_{\text{NNN}} > 0$, the threshold for the \mufix{} case shifts to $U^\mufix_c \sim 10.0$ or larger, seemingly independent of $t_{\text{NNN}}$ in the investigated range. In the \hfcnv{} case, the threshold varies from $U^\hfcnv_c \sim 7.4$ for $t_{\text{NNN}} = 0.05$ to $U^\hfcnv_c \sim 4.8$ for $t_{\text{NNN}} = 0.1$. For the \hfgen{} case, the threshold appears insensitive to next-nearest-neighbor hopping within the explored range.

For $t_{\text{NNN}} / t \geq 0.05$ in the \mufix{} case, universality is lost for all $U$ values, and the susceptibilities peak at $\mathbf{q}_\mathrm{max} \approx \frac{1}{3}(\mathbf{b}_1 \pm \mathbf{b}_2)$ for $U \geq 3.5$. In the \hfcnv{} case, universality is preserved for $U \gtrsim 4.8$, while for $U \leq 3.5$ and $t_{\text{NNN}} / t = 0.1$, many wavevectors compete. In the \hfgen{} case, universality is preserved for $U \gtrsim 7.4$ at $t_{\text{NNN}} / t = 0.05$ and for $U \gtrsim 4.8$ at $t_{\text{NNN}} / t = 0.1$. Thus, the behavior of $ U_c $ with respect to $ t_{\text{NNN}} $ is irregular.

\section{Detailed analysis: Kagome lattice}
\label{app:kagome}
Just as for the triangular lattice, we analyze specific heat $ c_v(T) $, density of states at the Fermi edge $ \rho_c(E_F, T) $, gap at the Fermi edge $ \Delta(E_F, T) $, and inverse participation ratio at the Fermi edge IPR($E_F, T$) on the kagome lattice. The plots of those quantities for the three filling cases are shown in Figs.~\ref{fig:N0_data_kagome}--\ref{fig:N1_data_kagome}. We provide a detailed discussion of each case below.

\subsection{Case \texorpdfstring{\mufix{}}{FCP}: Fixed chemical potentials, \texorpdfstring{$\mu_f = \mu_c = U/2$}{both equal to U/2}}
The peaks in the specific heat disappear completely or shift to very low temperatures ($T < 0.02$). No universal behavior is observed for $T \gtrsim 0.02$. For $U \gtrsim 8.7$, a peak at the lowest accessible temperatures appears, suggesting a possible quantum phase transition at zero temperature.

The IPR$(E_F)$ is similar to the triangular lattice \hfcnv{} case, with asymptotic values somewhat lower than those of the triangular lattice. Similar to the triangular \mufix{} and \hfgen{} cases, the asymptotic IPR values increase steadily by increasing the Coulomb interaction strength (from 0.01 at $ U = 1.0 $ to 0.16 at $ U = 10.0 $). These IPR values are slightly lower than those for the triangular lattice.

The behavior of $ \rho_c(E_F) $ for $ U \gtrsim 3.5 $ is qualitatively similar to that of the triangular \mufix{} case; changes are apparent only at lower temperatures, $ T < 0.02 $. The gap at the Fermi edge, $ \Delta_\mufix^\text{kagome} $ is initially close to $ U / 2 $ and drops to zero at $ T \sim 0.02 $.
The high-temperature value of the gap increases by increasing $ U $ (from 0.02 at $ U = 1.0 $ to 0.50 at $ U = 10.0 $).

The behavior of average occupations of f- and c-electrons, $ \bar{n}_f $ and $ \bar{n}_c $, is similar to the triangular lattice \mufix{} case.

Charge susceptibilities show very broad peaks in the range $T \in [0.02, 0.04]$ with no particular wavevector dominating. For $U \geq 8.7$, the peaks shift to $T < 0.02$, with $\mathbf{q}_\mathrm{max} = 0$, signaling a possible Mott transition. Universality is apparently lost.

\subsection{Case \texorpdfstring{\hfgen{}}{GHF}: Generalized half-filling, \texorpdfstring{$\bar{n}_f = 1/3$, $\bar{n}_c = 2/3$}{nf = 1/3, nc = 2/3}}
Instead of the sharp specific heat peaks of the triangular lattice \hfgen{} case, a very broad peak appears for $U \gtrsim 3.5$, with its maximum at $T \sim 0.03$.
The IPR$(E_F)$ is similar to the triangular \hfgen{} case with no perceivable peak at the transition temperature. The asymptotic IPR values increase steadily
by increasing the Coulomb interaction strength (from 0.03 at $ U = 1.0 $ to 0.1 at $ U = 10.0 $).

The behavior of $ \rho_c(E_F) $ for $ U \gtrsim 3.5 $ is qualitatively similar to that
of the triangular \hfgen{} case; changes are apparent only at lower temperatures, $ T < 0.02 $.

The gap at the Fermi edge, $ \Delta_\hfgen^{\text{kagome}} $, is initially less than $ U $ and drops toward zero at $ T \sim 0.02 $. The gap is larger than the kagome lattice \mufix{} case. The high-temperature value of the gap increases by increasing $ U $ (from 0.02 at $ U = 1.0 $ to 0.50 at $ U = 10.0 $).

Charge susceptibilities show sharp peaks with centers at $T \leq 0.02$, and no particular wavevector dominates, signifying a possible quantum phase transition at $T = 0$ with $\mathbf{q}_\mathrm{max} = 0$. No universality is observed.

\subsection{Case \texorpdfstring{\hfcnv{}}{CHF}: Conventional half-filling, \texorpdfstring{$\bar{n}_f = \bar{n}_c = 1/2$}{nf = nc = 1/2}}
The specific heat peaks are completely flattened for $U > 2.3$. A very broad peak appears for $U \gtrsim 3.5$. The IPR$(E_F)$ is similar to triangular \hfcnv{} case. The gap at the Fermi level is small, $\Delta^\text{kagome}_\hfcnv < \Delta^\text{kagome}_\hfgen$.

The behavior of IPR at the Fermi edge, IPR($ E_F $), is similar to the triangular lattice \hfcnv{} case in that the low-temperature peak (at $ T \sim 0.02 $) has a lower value than the asymptotic high-temperature value. The asymptotic IPR values increase steadily by increasing the Coulomb interaction strength (from 0.01 at $ U = 1.0 $ to 0.15 at $ U = 10.0 $).

The behavior of $ \rho_c(E_F) $ is qualitatively similar to that of the triangular \hfcnv{} case.

As in the triangular \hfcnv{} case, the gap at the Fermi level begins at a relatively low value $ \Delta_\hfcnv^\text{kagome} < \Delta_\hfgen^\text{kagome} $ for $ T \geq 0.02 $ and remains low for higher temperatures (from 0.02 at $ U = 1.0 $ to 0.50 at $ U = 10.0 $).

As in the other kagome cases, the susceptibilities show sharp peaks at $T \leq 0.02$ with $\mathbf{q}_\mathrm{max} = 0$, and no universality is observed.

\end{appendices}

\printbibliography

\end{document}